\documentclass[apj]{emulateapj}
\usepackage{epsfig}
\usepackage{amsmath}
\usepackage{amssymb}
\usepackage{natbib}
\usepackage{setspace}
\usepackage{graphicx}
\usepackage{array}

\newcommand{\degree}{\ensuremath{^\circ}}

\begin{document}
\title{Reconstructing the stellar mass distributions of galaxies using S$^4$G IRAC 3.6 and 4.5 $\mu \sc{m}$ images: I. Correcting for contamination by PAH, hot dust and intermediate-age stars}
\date{\today}
\author{Sharon E. Meidt\altaffilmark{1},
Eva Schinnerer\altaffilmark{1},  
Johan H. Knapen\altaffilmark{2,3}, Albert Bosma\altaffilmark{4}, E. Athanassoula\altaffilmark{4}, Kartik Sheth\altaffilmark{5,6,7}, 
Ronald J. Buta\altaffilmark{8}, 
Dennis Zaritsky\altaffilmark{9}, Eija Laurikainen\altaffilmark{10,11}, Debra Elmegreen\altaffilmark{12}, Bruce G. Elmegreen\altaffilmark{13}, Dimitri A. Gadotti\altaffilmark{14}, Heikki Salo\altaffilmark{10}, Michael Regan\altaffilmark{15}, Luis C. Ho\altaffilmark{16}, Barry F. Madore\altaffilmark{16}, Joannah L. Hinz\altaffilmark{9}, Ramin A. Skibba\altaffilmark{9}, Armando Gil de Paz\altaffilmark{17}, Juan-Carlos Mu\~noz-Mateos\altaffilmark{5}, Kar\'{i}n Men\'{e}ndez-Delmestre\altaffilmark{16}, Mark Seibert\altaffilmark{16}, Taehyun Kim\altaffilmark{5}, Trisha Mizusawa\altaffilmark{6,7}, Jarkko Laine \altaffilmark{10,11}, S\'{e}bastien Comer\'{o}n\altaffilmark{18} 
}

\altaffiltext{1}{Max-Planck-Institut f\"ur Astronomie / K\"{o}nigstuhl 17 D-69117 Heidelberg, Germany}
\altaffiltext{2}{Instituto de Astrofisica de Canarias, Spain}
\altaffiltext{3}{Departamento de Astrof\'\i sica, Universidad de La Laguna, Spain}
\altaffiltext{4}{Laboratoire d'Astrophysique de Marseille (LAM)}
\altaffiltext{5}{National Radio Astronomy Observatory}
\altaffiltext{6}{Spitzer Science Center}
\altaffiltext{7}{California Institute of Technology}
\altaffiltext{8}{University of Alabama}
\altaffiltext{9}{University of Arizona}
\altaffiltext{10}{University of Oulu, Finland}
\altaffiltext{11}{Finnish Centre for Astronomy with ESO (FINCA), University of Turku}
\altaffiltext{12}{Vassar College}
\altaffiltext{13}{IBM Research Division, T.J. Watson Research Center}
\altaffiltext{14}{European Southern Observatory}
\altaffiltext{15}{Space Telescope Science Institute}
\altaffiltext{16}{The Observatories of the Carnegie Institution for Science}
\altaffiltext{17}{Universidad Complutense Madrid}
\altaffiltext{18}{Korea Astronomy and Space Science Institute}
%%%%%%%%%%%%%%%%%%%%%%%%%%%%%%\doublespacing
\begin{abstract}
With the aim of constructing accurate 2D maps of the stellar mass distribution in nearby galaxies from S$^4$G 3.6 and 4.5 $\mu m$ images, we report on the separation of the light from old stars from the emission contributed by contaminants.  Results for a small sample of six disk galaxies (NGC 1566, NGC 2976, NGC 3031, NGC 3184, NGC 4321, and NGC 5194) with a range of morphological properties, dust contents and star formation histories are presented to demonstrate our approach.  To isolate the old stellar light from contaminant emission (e.g. hot dust and the 3.3 $\mu  m$ PAH feature) in the IRAC 3.6 and 4.5 $\mu m$ bands we use an Independent Component Analysis (ICA) technique designed to separate statistically independent source distributions, maximizing the distinction in the [3.6]-[4.5] colors of the sources.  The technique also removes emission from evolved red objects with a low mass-to-light ratio, such as asymptotic giant branch (AGB) and red supergiant (RSG) stars, revealing maps of the underlying old distribution of light with [3.6]-[4.5] colors consistent with the colors of K and M giants.  The contaminants are studied by comparison to the non-stellar emission imaged at 8$\mu m$, which is dominated by the broad PAH feature.  
Using the measured 3.6$\mu m$/8$\mu m$ ratio to select individual contaminants, we find that hot dust and PAH together contribute between $\sim$5-15\% to the integrated light at 3.6 $\mu m$, while light from regions dominated by intermediate-age (AGB and
RSG) stars accounts for only 1-5\%.  Locally, however, the contribution from either contaminant can reach much higher levels; dust contributes on average 22\% to the emission in star-forming regions throughout the sample, while intermediate age-stars contribute upwards of 50\% in localized knots.   
The removal of these contaminants with ICA leaves maps of the old stellar disk that retain a high degree of structural information and are ideally suited for tracing stellar mass, as will be the focus in a companion paper.  
\end{abstract}
\section{Introduction\label{sec:intro}}
Accurate maps of the stellar mass distribution in nearby galaxies are essential for charting the dynamical influence of stellar structures over time.  They provide a key perspective on the evolution in the amount and distribution of baryons from the high redshift to the local Universe, both as the result of internal, secular processes (e.g. \citealt{kormKenn}; \citealt{elm07}; \citealt{zhang}; \citealt{haan}; \citealt{foyle}; \citealt{elms09}; \citealt{elmelm10}), and given `bottom up' assembly and external (heirarchical accretion) events within the context of CDM cosmology ( 
e.g., \citealt{balcells} and \citealt{courteau};  \citealt{florido}; and see \citealt{navw}; \citealt{somerville}; \citealt{governato}; \citealt{guo}).

Yet, stellar mass estimation as a field is still maturing.  NIR bands are often exploited as optimal windows on the old stars that dominate the baryonic mass in galaxies, with images at these wavelengths serving as good relative proxies for stellar mass (\citealt{ee84}; \citealt{rixzaritsky}; \citealt{grosbol}).  But estimates incorporating dependencies on star formation history, metallicity, and especially the age of the stellar population (\citealt{rixrieke}; \citealt{rhoads}), as well as dust extinction and reddening, have only relatively recently emerged, following the demonstrated correlation between optical colors and the mass-to-light ratio in stellar population synthesis models (\citealt{bdJ}; \citealt{z09}).  

The Spitzer Survey of Stellar Structure in Galaxies (S$^{4}$G; \citealt{sheth}) provides an un-paralleled inventory of stellar mass and structure in nearby galaxies, with deep imaging 
at 3.6 $\mu m$ and 4.5 $\mu m$ (reaching $\mu_{3.6\mu m}(AB)(1\sigma)$$\sim$27 mag arcsec$^{-2}$) where the light mainly traces the oldest stars. 
The sample of S$^4$G galaxies therefore stands to significantly contribute to our understanding of the evolution of stellar mass and structural properties.  

For the purposes of accurately mapping the radial distributions of mass in galaxies, and in particular later Hubble types, the situation remains less than ideal.  Although M and K giants should in fact be the foremost contributors to the flux in these two shortest IRAC wavebands, 
by now several other sources alongside the oldest stellar light are well-discerned at IRAC wavelengths in star-forming galaxies.   
The finite instrumental bandwidth introduces additional features (i.e. the 3.3 $\mu m$ PAH emission feature in the 3.6 $\mu m$ band and the PAH continuum at 4.5 $\mu m$; \citealt{flagey}), accessory non-stellar continuum emission (i.e. from hot dust) appears at these wavelengths, and lower-M/L asymptotic giant branch (AGB) and red supergiant (RSG) stars can also make significant contributions to the emission.   

Depending on physical conditions in the disk, the degree of active compensation for these sources varies from galaxy to galaxy, and different investigators propose different compensation schemes.  Total mass estimates should be only modestly affected by the presence of contaminants, which contribute a few percent to the integrated flux.  But 1D and/or 2D representations of either the relative or absolute stellar mass estimated from 3.6 $\mu m$ images can be greatly influenced by the pronounced local contribution made by these sources. These three main contaminants can lead to over-estimation in stellar mass maps and therein complicate measurements of structural diagnostics.  

Several schemes for the identification (if not correction) of these features in IRAC images have been developed (\citealt{kendall}; \citealt{hunter}; \citealt{bolatto}; \citealt{deblok}), but only few with the aim of deriving maps of the mass distribution in nearby galaxies.  Given that PAH and hot dust contaminants appear in star-forming (high-density) spiral arm regions, distinct components fits to the stellar emission there are prone to inaccuracy and so are typically omitted in favor of an overall azimuthally-averaged contribution in surface brightness models (\citealt{kendall} and \citealt{leroy}).  While these methods can often supply a contamination-free view of the old stellar disk, it is not immediately clear that they can be successfully extended to a large data set like S$^4$G, which samples throughout the Hubble sequence, over a large range of dust contents, gas fractions, and star formation histories.  
Furthermore, the auxiliary data employed by these methods (e.g. 8 $\mu m$, UV, optical images) will not be uniformly available for the full S$^4$G sample of 2,331 galaxies.  A correction for contaminant emission at 3.6 $\mu m$ and/or 4.5 $\mu m$ that is self-reliant, and can assure uniformity throughout the sample, is therefore highly desirable.

Corrections for non-stellar emission are especially critical for mass estimates, since this emission is not incorporated into SED libraries at the core of now-standard techniques  
for estimating the stellar M/L (e.g. \citealt{bdJ}; \citealt{z09}).   Tight constraints on the stellar M/L are possible using optical or NIR colors that leverage metallicity, age and SFH variations in the underlying stellar population.  Recently, Zibetti et al. (2009) have shown that a second, optical/NIR color can significantly boost the predictive power of the color-M/L relations first assembled by \citet{bdJ}.  With color maps replacing global colors, plus a prescription to account for the effects of dust extinction/reddening, local estimates of the M/L in NIR bands can be obtained with high accuracy.  The few existing 2D maps of the stellar mass distribution derived in this way show great promise, and reveal a unique glimpse of the changing appearance of stellar structures between light and mass.   Paired with similar information, S$^4$G images free of emission from contaminants should be readily convertible into high-quality mass maps, as considered in Paper II of this series \citep{meidtII}.  

To complement S$^4$G science, we therefore introduce a technique that removes contaminants from S$^4$G images to reveal an underlying uniformly old distribution of stellar light, with [3.6]-[4.5] colors that are consistent with those of M and K giants (\citealt{pahre2}; \citealt{hunter}).   

The Independent Component Analysis (ICA) technique used to pursue these ends (described in the next section) in principle allows much more information in individual structural components, like spiral arms, to be maintained in contaminant-corrected mass maps than is supplied by other commonly-employed mass estimation methods; models of the 3.6 $\mu m$ and/or 4.5 $\mu m$ emission (wherein the appearance of contaminants is reduced on azimuthal averaging), have so far provided the basis for most mass estimates at these wavelengths (e.g. \citealt{deblok}; \citealt{leroy}).   Our maps of the old stellar light, with full structural information, are ideal relative mass tracers for optimal measurement of bar strength and ellipticity (e.g. \citealt{menendezdelmestre}), spiral arm-interarm contrasts and relative Fourier amplitudes and phases (e.g. \citealt{laurikainen}; \citealt{salo2010}; \citealt{elmegreen2011}).  Likewise, we anticipate that these maps can be converted into high-quality absolute mass maps relevant for deriving the underlying stellar potential, measuring bar and/or spiral arm torques (e.g. \citealt{stark}; \citealt{zaritskyLo}; \citealt{block}; \citealt{haan}; \citealt{salo2010}) and mass flux/gas inflow (e.g. \citealt{quillen}; \citealt{zhang}), and for constraining the baryonic contribution to observed rotation curves (e.g. \citealt{deblok}).  

In $\S$ \ref{sec:icaapp} we describe our approach (motivated in $\S$ \ref{sec:motivation}), with particular emphasis on the use of 3.6 and 4.5 $\mu m$ S$^4$G data and data products ($\S\S$ \ref{sec:icause} and \ref{sec:apost}).  Results from the application of ICA to a small sample of galaxies (described in $\S$ \ref{sec:data}) using S$^4$G archival images processed through the S$^4$G reduction pipeline (\citealt{sheth}) are presented in $\S$ \ref{sec:sources}.  These ICA sources are identified as either old stars or contaminants, where the latter is defined as all sources of emission apart from the oldest  
stars: PAH, hot dust, and intermediate-age red supergiant (RSG) and asymptotic giant branch (AGB) stars. 

In $\S$ \ref{sec:contams} we focus on the contaminant emission and relate the intensity and spatial information contained in our ICA source maps to archival IRAC channel 4 images, which are dominated by the broad 7.7 $\mu m$ PAH feature.  This comparison reveals the detection of both the 3.3 $\mu m$ PAH feature and hot dust in 3.6 $\mu m$ images with ICA ($\S$ \ref{sec:dust}).  It also allows for the identification of a second, non-dust contaminant at 3.6 and 4.5 $\mu m$, namely the bright but low M/L red objects contributing to the emission at these wavelengths ($\S$ \ref{sec:rsgagb}).  

The collective removal of these contaminants with ICA reveals maps of the underlying old stellar light that should be readily convertable into mass maps.  In a forthcoming paper we will examine our ICA maps of the old stellar light in detail, and compare different techniques for assigning the mass-to-light ratio at 3.6 $\mu m$.  Here we include a preliminary assessment of the quality of these maps (as well as of the contaminant correction, itself) by comparing the old stellar emission at 3.6 $\mu m$ with other NIR tracers.  Finally in $\S$ \ref{sec:conclusions} we summarize our findings and describe avenues for future work with the ICA-identified sources.  
\section{On the application of ICA for correcting S$^4$G images\label{sec:s4gica}}
\subsection{Motivation\label{sec:motivation}}
Mass maps for much of the S$^4$G sample will necessarily rely on information from only the two shortest IRAC wavebands available during the Spitzer warm mission.  Given this constraint, the task of removing contaminants from these images might involve some sort of scaling and subtraction of the two measurements based on an (S$^4$G-defined) empirical recipe.\footnote{A recipe for scaling and subtracting the 3.6 and 4.5 $\mu m$ images to remove contaminants could be established, e.g., using a calibration to 8 $\mu m$ data.  Note, however, that simply subtracting a scaled version of the 4.5 $\mu m$ image from the 3.6 $\mu m$ image cannot account for the fact that dust and stellar SEDs have different shapes between these two bands, and a single scaling is not physically realistic throughout the entire galaxy. }  On the other hand, the constraints also suggest an approach that is similarly-premised and yet less susceptible to empirical interpolation-uncertainties across the sample, i.e. as would be introduced by assuming a uniform spectral shape for the stellar and dust SEDs for all galaxies between 3.6 and 4.5 $\mu m$.  

Specifically, let us consider the emission at frequency $\nu$ as the superposition of the emission from $N$ different sources (stellar and non-stellar) such that, for $P$ samples (viz. pixels) in $M$ frequency channels,
\begin{equation}
\mathbf{x}=\mathbf{A}\cdot\mathbf{s}\label{eq:ica}
\end{equation}
where $\textbf{x}$ is the $M$$\times$$P$ measurement set, $\textbf{A}$ is an $M$$\times$$N$ mixing matrix containing the coefficients of the combination and the $N$$\times$$P$ matrix $\textbf{s}$ represents the source signals, under the assumption that the sources $\textbf{S}$ are separable functions of location and frequency (i.e. $s_j$=$S_j(\textbf{r})$ and $A_{i,j}\propto S_{j}(\nu)$). 

In this way, we can extract measurements of the source signals $s_j$ (each representing either contaminants or emission from the true old stellar disk) as some linear combination of input channels $x_i$ (i.e. IRAC channel 1 and 2 images at 3.6 and 4.5 $\mu m$).  But rather than assume a particular set of separation coefficients $A^{-1}_{ij}$ (given some empirical recipe), we can solve for these simultaneously, letting source extraction become a problem of so-called blind source separation (BSS).

Several techniques have been proposed to reliably solve problems of this type, 
including Principal Component Analysis (PCA) and its increasingly popular affiliate Independent Component Analysis (ICA).   Both extract source measurements via transformation of the input data, but whereas the former imposes only that the sources be uncorrelated (i.e. they have zero covariance), the latter maximizes their statistical independence.  This grants greater uniqueness to ICA solutions, promising improved descriptions of sources that are the result of distinct physical processes.  
As demonstrated in $\S$ \ref{sec:sources}, ICA decomposes the images at 3.6 and 4.5 $\mu m$ each into two other images, tracing either the old stars or the contaminant sources of emission, by maximizing the distinction in the [3.6]-[4.5] color of these components.

Here we employ the FastICA realization of noise-free ICA developed by \cite{hica} and \cite{hoica}.\footnote{We use an implementation of FastICA from within the IT++ library, publicly available for download at http://itpp.sourceforge.net}  This powerfully efficient method uses a negentropy measure of non-``Gaussianity'' to define the statistical independence of the sources.  By the central limit theorem, a mixture of independent variables is more Gaussian than either of the two alone, and so the requirement of variable-independence can be replaced with the equivalent requirement for less ``Gaussianity" in these variables.

Thorough descriptions of the method can be found, e.g., in the analysis of multi-channel COBE and WMAP observations of the CMB (\citealt{maino1}; \citealt{maino2}).   To summarize briefly here, the separation of non-gaussian sources is achieved through linear combinations of the input data at different frequencies (e.g. $\textbf{y}$=$\textbf{W}\cdot\textbf{x}$ with $\textbf{y}$$\sim$$\textbf{s}$ and $\textbf{W}$$\sim$$\textbf{A}^{-1}$) corresponding to maxima of the negentropy.  The negentropy, itself, is approximated by a non-linear re-mapping of the input data according to one of several functions $g$ of the data's iteratively updated principal components $u$, the choice of which controls the form of the identified source distributions.  In this work, we employ the function $g(u)\propto exp(u^2)$ (see e.g. \citealt{hoica}), which is thought to be well-suited to cases where the independent components are highly super-Gaussian (\citealt{hica}; as opposed to the kurtosis function, which is more sensitive to outliers in the source distributions) and which \citet{maino2} find achieves the best recovery of CMB frequency scaling.

By the nature of the method, sources determined with ICA can take on negative values since they are assumed to have statistically independent distributions with zero mean.  
Nevertheless, we favor this method to a third technique, Non-negative Matrix Factorization (NMF), which, despite promising physically realistic solutions (given that sources should be positive), is more prone to non-unique solutions whereby true source characterization is less immediately obvious.  

Although sparseness constraints on NMF (e.g. NMFSC; \citealt{hoyer}) have been shown to improve the uniqueness of solutions in some applications, neither the sparseness of the sources nor their `mixing' coefficients can be specified individually.  For our purposes, then, we find that any such separation of sources is still less satisfactory than that which can be achieved with ICA; we expect at least one source (i.e. emission from old stars) to appear both throughout the disk (in each pixel) and in the two wavebands.  
Any source negativity imposed by ICA is furthermore largely relegated to noise, which we find predominantly hosted by one source map (see eq. \ref{eq:noise} below).  

\begin{table*}
\begin{center}
\caption{Sample\label{tab-params}}
{\small 
\begin{tabular}{rccccccccc}
\tableline\tableline

 Galaxy&RA&DEC&Classification&$L_{IR}/L_{opt}$\footnote{Taken from \citet{kennSings}}&8/24$\mu m$\footnote{Calculated using the integrated fluxes measured by \citet{munozmateos}}&D (Mpc)&$R_{edge}$(')&$\epsilon_{edge}$&PA$_{edge}$($\degree$)\\
\tableline
NGC 1566&04$^h$20$^m$00.4$^s$ &-54$\degree$56'16''
&(R1Õ)SAB(s)b&0.49&0.69&17&4.5&0.996&4\\
NGC 2976&09$^h$47$^m$15.4$^s$ &+67$\degree$54'59''
&SAB(s:)d&0.49&1.09&3.6&3.75&0.599&54\\
NGC 3031&09$^h$55$^m$33.2$^s$ &+69$\degree$03'55''
&SA(s)ab\footnote{RC3 Classification \citep{rc3}}&0.08&0.66&3.6&11.12&0.55&70\\
NGC 3184&10$^h$18$^m$16.8$^s$ &+41$\degree$25'27''
&SA(rs)bc&0.99&0.90&8.6&3.25&0.99&2\\
NGC 4321&12$^h$22$^m$54.8$^s$ &+15$\degree$49'19''
&SAB(rs,nr)bc&0.73&0.81&18&5.62&0.83&68\\
NGC 5194&13$^h$29$^m$55.7$^s$ &+47$\degree$13'53''
&SAB(rs,nr)bc&0.6&0.82&8.4&5.62&0.65&90\\
\tableline
\end{tabular}
}
\tablecomments{The parameters defining the edge of the disk (last three columns) are chosen by eye to encompass the bulk of the disk emission at 3.6 $\mu m$.  Classifications are from \citet{buta10}, unless otherwise noted, and distances are taken from \citet{munozmateos1}.}
\end{center}
\end{table*}
\subsection{Application to S$^4$G images\label{sec:icaapp}}
Given two input samples (here the two images at 3.6 and 4.5 $\mu  m$) the method supplies at most two independent sources (i.e. $M$=$N$ for \textbf{A} invertible), imposing a strict limit on the number of contaminants we can detect, in addition to emission from the old stellar disk.   As discussed in $\S$ \ref{sec:multidists}, however, we find that two input images may be sufficient for the binary distinction between old stars and contaminants, given that the latter tend to be detected together as a single second source.  When this second source is removed from the first source, however, artifacts are introduced, as discussed in $\S$ \ref{sec:imperfections}.

With ICA measurements of the two sources, then, along with the 2$\times$2 separation matrix $\textbf{W}\simeq\textbf{A}^{-1}$, the components $j$ at a given pixel in each image $i$ can be reconstructed as $X_{ij}=W_{ij}^{-1}s_j$ (without implicit sum over $j$).  
(Note that for any BSS technique there is a scaling ambiguity between the mixing and sources, $A_{ij}$ and $s_j$, but the contribution of each source to the total flux can be written as A$_{ij}$s$_j$ without loss of information.)  By this definition, the flux of source $j$ in image $i$=1 at 3.6$\mu m$ is related to its flux in image $i$=2 at 4.5 $\mu m$ as $W^{-1}_{3.6,j}/W^{-1}_{4.5,j}$.  So, taking source $s_1$ as the old stars, for example, this relation defines an effective corrected global [3.6]-[4.5] color for the s1 component according to the ratio $A_{3.6,1}/A_{4.5,1}$.

Multiple realizations of the sources and their colors can be obtained by seeding the ICA calculation with a non-random initial guess for the mixing coefficients with controlled variation.  Our seeds conservatively suppose that the flux at 3.6 and 4.5 $\mu m$ is split between the stars and the dust such that these components have colors -0.2$<$[3.6]-[4.5]$_{stars}$$<$0 and 0$<$[3.6]-[4.5]$_{dust}$$<$1.5, respectively. The choice of seed is arbitrary in the sense that the colors of the calculated components quickly move away from the initial values and approach the final converged values, but this step allows us to define a measure of the uncertainty in the final solution: we take the average and rms variation of the colors returned with our fixed number of seeds and then adopt the average to reconstruct the quantity A$_{ij}$s$_j$ for each source at the two wavelengths.

\subsection{Comments on ICA with S$^4$G Data Products\label{sec:icause}}
To best utilize ICA for source identification in this particular application requires
that the emission intrinsic to the galaxy can be defined in the input images.  This prevents the ICA-identified distributions from being skewed, or biased, representations of the true distributions of emission sources.  
Measurements of the shape and location of the effective edge of the old stellar disk, such as measured by S$^4$G pipeline III (by S$^4$G convention at the 26.5 magnitude/arcsec$^2$ isophote; \citealt{sheth}),
are therefore a crucial input for this technique. The parameters of the edge of the disk listed in Table \ref{tab-params} were chosen by eye to enclose the disk emission according to the S$^4$G definition.  Variation by $\sim$20\% in these values has negligible effects on the results, whereas significant differences arise when, e.g. the radius is defined to be half as large. 

Masks of bad pixels (e.g. S$^4$G masks prepared with SExtractor) also provide a convenient way to avoid features in the input images, including field stars and background galaxies, that are clearly not inherent to the galaxy.
However, we do not use specially designed masks in the ICA analysis that follows.  In testing we find that this choice  has very little effect on the results of the method for this sample of galaxies, despite the risk of bias (i.e. accurate source extraction with ICA requires that the intrinsic shape of the underlying source distribution be correctly identified); for the six galaxies here, the identified distributions vary little in cases with and without field stars masked.  Such masking may be necessary in other cases with stronger foreground contamination, however, and should add decided value at a stage of analysis beyond the application of ICA (e.g. GALFIT decompositions of the surface brightness distribution), for accurate estimation of the total flux and mass of the old stellar population.  Other features with the potential to bias the results of the analysis, such as observational artifacts, are identified and excluded at a later stage, as described in $\S$ \ref{sec:apost}.

Since source separation with ICA leverages color information, the small differences in the 3.6 and 4.5 $\mu m$ PSFs are a potential influence on accuracy of the solutions.  Here we have used the convolution kernel calculated by \citet{ganiano} to match the 3.6 and 4.5 $\mu m$ images.  The results in this case are very similar to those obtained using unmatched images, though, with less than 0.01 mag difference to the final source colors.  The choice of kernel (i.e. substituting the \citealt{gordon} kernel) or input PSFs (using the S$^4$G ``super''-PSFs; \citealt{sheth}) also does not significantly impact the results. This is to be expected given that the PSFs for the 3.6 and 4.5 $\mu m$ bands are comparably sized (although highly structured in the wings; \citealt{reach} and \citealt{ganiano}).  Any effects due to lingering PSF mismatch can also be minimized during the \emph{a posteriori} exclusion described in $\S$ \ref{sec:apost}. For application to the full S4G sample of extended sources the necessity of PSF-matching is therefore unclear, but care may be warranted in select cases (i.e. bright or saturated point sources in the field). 

Prior to source separation with ICA, the sky background is subtracted from the 3.6 and 4.5 $\mu m$ images for each galaxy.  Here we define the sky level in either band as the average of pixels with signal-to-noise (S/N) $<$2 beyond the edge of the disk.  We note that the adopted sky level has very little influence on the identified sources (except for a uniform shift in flux level), and does not affect the scaling between them.  Colors listed in Table \ref{tab-colors} are therefore only subject to photometric errors (defined as by \citealt{salo}; plus $\sim$20\% for the sky error), which typically range from $\sim$ 0.03 mag at $\mu_{3.6}$=18 mag arcsec$^{-2}$ to 0.2 mag at $\mu_{3.6}$=21 mag arcsec$^{-2}$ at 3.6 $\mu m$ in a given pixel.  

According to eq. (\ref{eq:ica}), the flux errors in both sets of output source images (one pair at each wavelength) can be expressed as a combination of the errors in the input images, $\sigma_1$ and $\sigma_2$.  At 3.6 $\mu m$ (band 1), for example,
\begin{eqnarray}
\sigma_{11}&=&\frac{1}{1-10^{-0.4(c1-c2)}}(\sigma_1+z10^{-0.4c2}\sigma_2) \nonumber \\
\textrm{and} \nonumber \\
\sigma_{12}&=&\frac{10^{-0.4(c1-c2)}}{1-10^{-0.4(c1-c2)}}(\sigma_1+z10^{-0.4c1}\sigma_2)
\label{eq:noise}
\end{eqnarray}
in the first and second source images, respectively.  Here, $z$ is the ratio of the zeropoints in each measurement band, and $c_1$ and $c_2$ are the colors for each source (e.g. as listed in Table \ref{tab-colors}).  The noise in the final source images at the second wavelength (band 2), $\sigma_{21}$ and $\sigma_{22}$, are equivalent to $\sigma_{11}$ and $\sigma_{12}$ with only a scaling determined by the color of the source.   In the following, photometric errors $\sigma_m$ refer to these errors on the fluxes measured in a given source image $m$ (dropping the waveband index).
\subsection{A posteriori exclusion\label{sec:apost}}
Both source maps exhibit typically noisy fluctuations beyond the bulk of the emission, where the signal-to-noise  in the original maps is low.  Only rarely where the S/N is otherwise high will negative values persist, primarily in map s2.  This `ghosting' is symptomatic of the incompatibility of the true sources with the set of only two that we calculate here with ICA (as identified also by Maino et al. 2003).  
This could be intrinsic to sources in the disk (e.g. stellar distributions with distinct old and young components), but here it occurs mostly as the result of observational artifacts, i.e. in the presence of a saturated nucleus or due to PSF mismatch around bright central sources.  
In such cases we recalculate solutions with any such pixels masked in the input measurement sets.  For saturated nuclei a circular mask is chosen by eye around the central region, while in all other cases masking occurs where negative-valued pixels exceed $\vert 10\sigma_m\vert$ (i.e. where $\vert S/N\vert$$>$10) in either source image at 3.6 $\mu m$.  If the number of pixels in the two recalculated source maps is in total reduced, these solutions replace the original set.  Flux information from the masked region is then passed untouched into either one of the two component maps, depending on which of the first source solutions defines the \emph{a posteriori} mask.  In cases where negative values initially appear in s2, for example, the new map s1 regains the original flux in the masked region, while pixels in the new s2 are there set to zero.   Note that, with \emph{a posteriori} exclusion, the final maps reflect a combination of the uniform ICA source and a subset of pixels with flux and color information from the original uncorrected image.  

We emphasize that these instances are limited (here it occurs in the central regions of NGC 5194 and M81 and at the location of the saturated nucleus in NGC 1566, covering less than 1\% of the disk area in each case).  Negative-valued pixels are otherwise mostly all within the 1$\sigma_m$ uncertainty for each pixel defined by Poisson statistics, and so can be considered equivalent to a non-measurement.  

In an additional post-processing step, these negative pixels are excluded from final ICA maps, together with all pixels below a 5$\sigma_m$ flux threshold, where $\sigma_m$ for each source image is calculated from the noise in the original 3.6 and 4.5 $\mu m$ images according to eq. \ref{eq:noise}.  
Pixels in either map with fluxes below the cut-off are then replaced with zero (or blank), while pixels at those positions in the companion map are set to the value measured in the original image (i.e. flux in an image below the threshold is added back into the companion map).  

This thresholding produces final source images that more realistically represent the expected spatial distribution of contaminants; in the absence of real flux from a contaminant, ICA s2 `contaminant' images are otherwise filled with noise over a large area inside the disk edge.  The final companion old stellar images in this way also better portray `contaminant-free' regions.  With the exclusion of pixels in the ICA contaminant map below 5$\sigma_{2}$ at 3.6 $\mu m$, for instance, pixels in the old stellar map at the positions of these blanked pixels are returned to their values in the original, uncorrected maps.  (In all cases, the 5$\sigma_2$ flux threshold is by far more exclusive in the contaminant map than the 5$\sigma_1$ threshold is in the map of the old stellar light,  so that flux is more often reintroduced to the old stellar map, over a larger fraction of the disk.)  

This \emph{a posteriori} adjustment to the source fluxes also varies the ratio of the sources between 3.6 and 4.5 $\micron$ from the global values designated by ICA.  For instance, the [3.6]-[4.5] colors in the contaminant free regions of the old stellar maps take on the values originally observed in the uncorrected images. 
According to our adopted 5$\sigma_2$ flux threshold, this change to the old stellar color is on the order of the 1$\sigma$ photometric error in the two bands, given the typically $\sim$20\% difference in errors between the 3.6 and 4.5 $\mu m$ bands.   (From equation \ref{eq:noise}, the change in the old stellar map color with the reintroduction of flux from the contaminant map below the 5$\sigma_2$ threshold is at most equal to the difference between the 5$\sigma$ errors in the two bands, i.e. 5$\sigma_{3.6}$-5$\sigma_{4.5}$ as defined in $\S$ \ref{sec:icause}.  Higher thresholds can be used at the expense of the [3.6]-[4.5] color.)   In what follows we take the ICA [3.6]-[4.5] color as representative of the true old stellar disk, and assert that this ICA color is realistic to within 1$\sigma$ throughout the entire final map of the primary, old stellar source.  Note that, without ICA, the [3.6]-[4.5] color can be difficult to reliably measure in the outer portions of the disk, given the low S/N there.  However, with ICA we gain a measure of the [3.6]-[4.5] color that is good to 1$\sigma$, according to our thresholding, given the use of information from throughout the rest of the disk.

\section{The S$^4$G dataset}
\label{sec:data}
The analysis presented in this work is based on 3.6 $\mu m$ and 4.5 $\mu m$ S$^4$G archival images which have been processed through the S$^4$G reduction pipeline (first column in Figure \ref{fig-4321}), as described by \citet{sheth}.  The subset of galaxies considered here, NGC 1566, NGC 2976, NGC 3031 (M81), NGC 3184, NGC 4321 and NGC 5194 (M51) (see Table \ref{tab-params}), were chosen for their range of morphological properties, dust contents, and star formation histories from SINGS (\citealt{kennSings} and see, e.g., \citealt{draine07}, \citealt{calzetti}, \citealt{bendo}, \citealt{munozmateos}, \citealt{moustakas}).  In addition to a fairly wide range in FIR-to-optical flux ratio (Table \ref{tab-params})--which is correlated with dust optical depth, dust temperature and inclination \citep{kennSings}--these galaxies span a range in nuclear activity: Seyfert (NGC 1566, M51), LINER (NGC 3031, NGC 4321), and starburst (NGC 2976, NGC 3184) \citep{smith}.  They also show a diversity in PAH emission relative to hot dust, as measured by the ratio 8/24 $\mu m$ (Table \ref{tab-params}), typical for normal star-forming galaxies \citep{dale}.  Finally, typical of SFHs for Hubble types Sa-Sd this sample of six galaxies covers 2.8$<$FUV-[3.6]$<$4.5, with redder colors indicating lower current-to-past star formation activity \citep{munozmateos}.  
Although few in number, this subset is expected to be diverse enough to demonstrate the projected effectiveness of the ICA technique over a range of Hubble Types.  

We compile a set of comparison data for these galaxies from SINGS (namely IRAC 8 $\mu m$ and MIPS 24$\mu m$ images; see \citealt{kennSings}) and 2MASS (here, primarily H-band images; see \citealt{jarrett}).  Additional processing of these images will be described when referred to in upcoming sections.

Unless otherwise specified, all images are presented in units of MJy/sr.  For conversion to mag$/$arcsec$^2$ relative to Vega, we use the zeropoint fluxes 280.9, 179.7 and 64.13 Jy at 3.6, 4.5 and 8$\mu m$, respectively \citep{reach}, and the 2MASS H-band magnitude zeropoint provided in the image header \citep{jarrett}.   Surface brightnesses in units of $L_{\odot} /pc^2$ are also derived using the value of the absolute magnitude of the sun in the IRAC 3.6 $\mu m$ band, $M_{\odot}^{3.6}$= 3.24 mag, calculated by \citet{oh}.  
\section{Maps of the sources identified with ICA}
\label{sec:sources}
\begin{figure*}[p] 
\begin{flushleft}
\begin{tabular}{m{.05in}m{1.25in}m{1.25in}m{1.25in}m{1.25in}m{1.25in}}
& \footnotesize \centering 3.6 $\mu m$& \footnotesize \centering [3.6]-[4.5]& \footnotesize \centering s1-ICA & \footnotesize \centering s2-ICA & \footnotesize \centering 8$\mu m$ \tabularnewline
%&&&&& \\
\rotatebox{90}{\footnotesize NGC 1566}&  \includegraphics[width=1\linewidth]{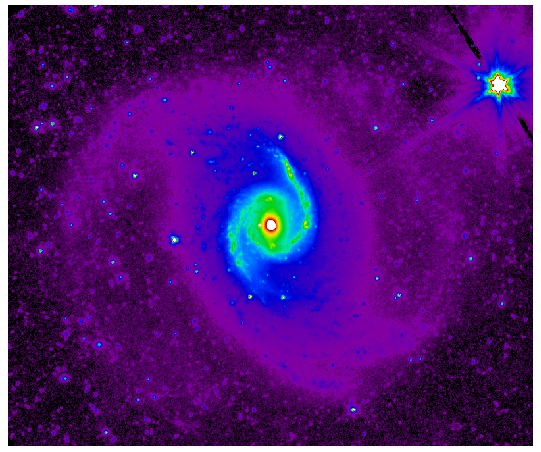}& \includegraphics[width=1\linewidth]{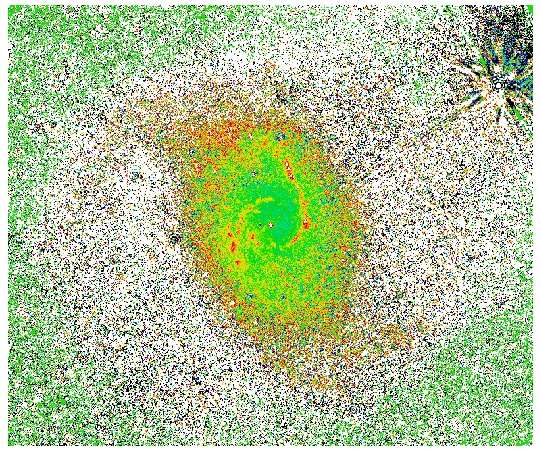}&
\includegraphics[width=1\linewidth]{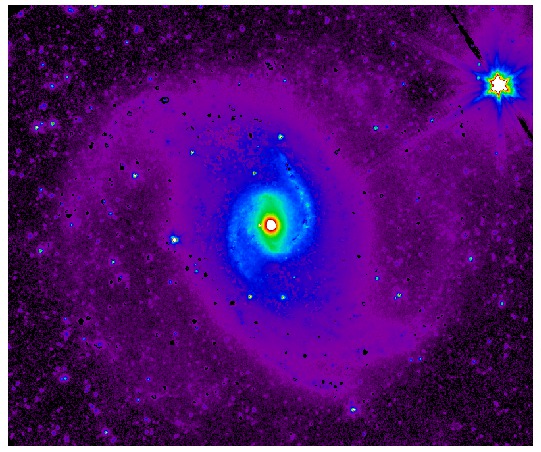}& \includegraphics[width=1\linewidth]{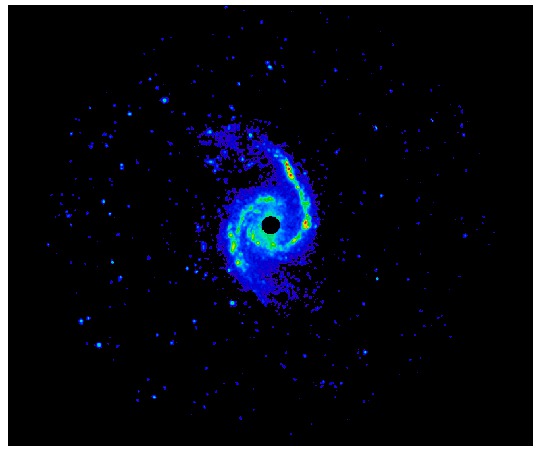}&
\includegraphics[width=1\linewidth]{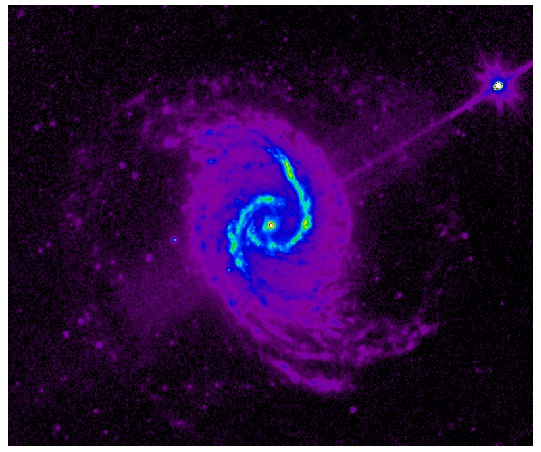}\\
\rotatebox{90}{\footnotesize NGC 2976}&  \includegraphics[width=1\linewidth]{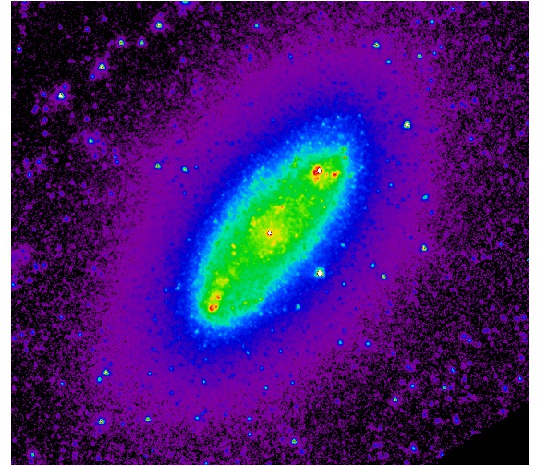}& \includegraphics[width=1\linewidth]{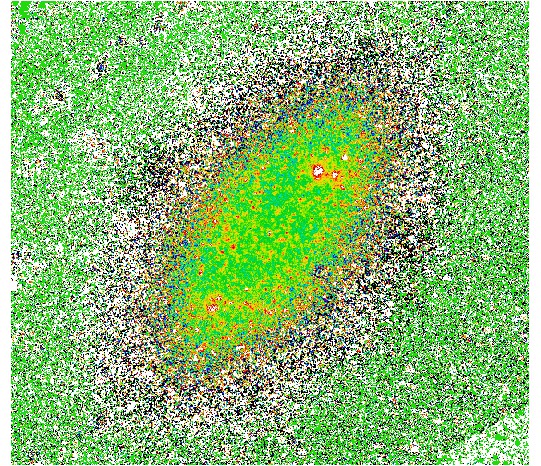}&
\includegraphics[width=1\linewidth]{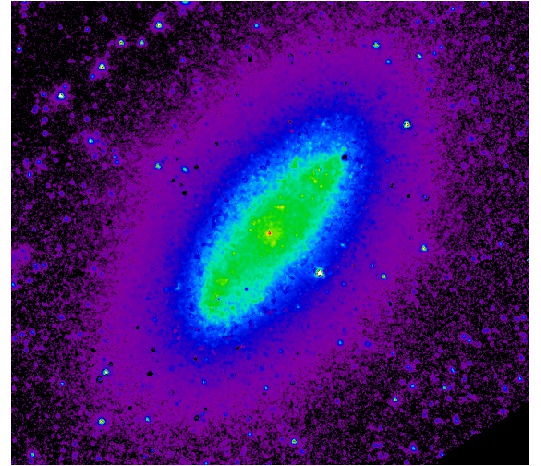}& \includegraphics[width=1\linewidth]{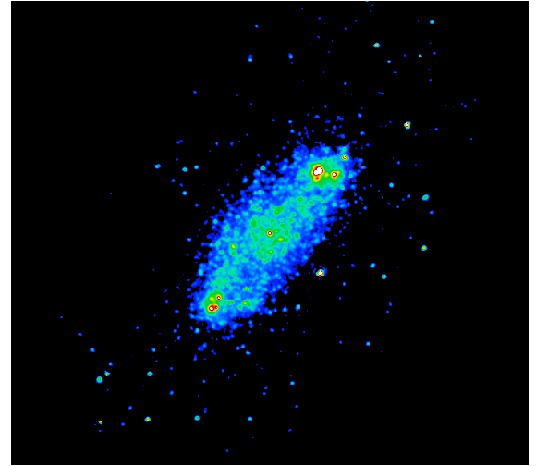}&
\includegraphics[width=1\linewidth]{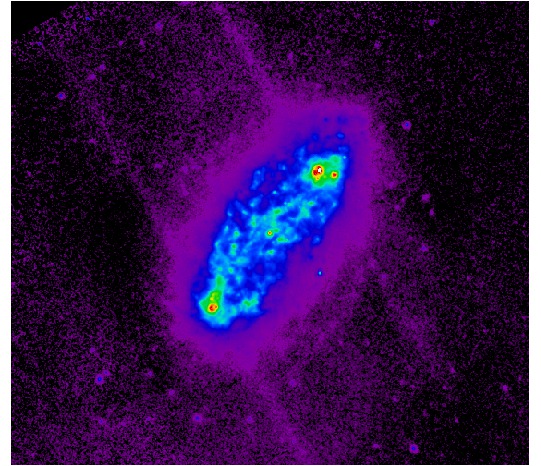}\\
\rotatebox{90}{\footnotesize NGC 3031} & \includegraphics[width=1\linewidth]{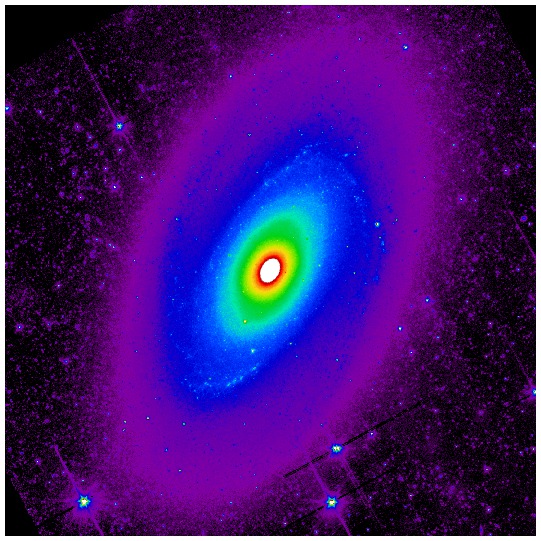}& \includegraphics[width=1\linewidth]{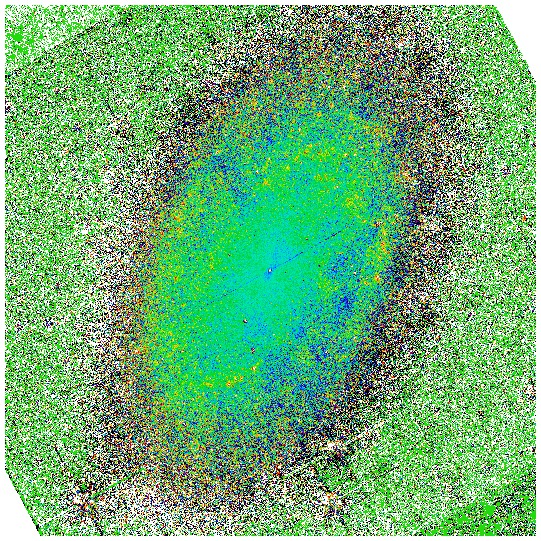}&
\includegraphics[width=1\linewidth]{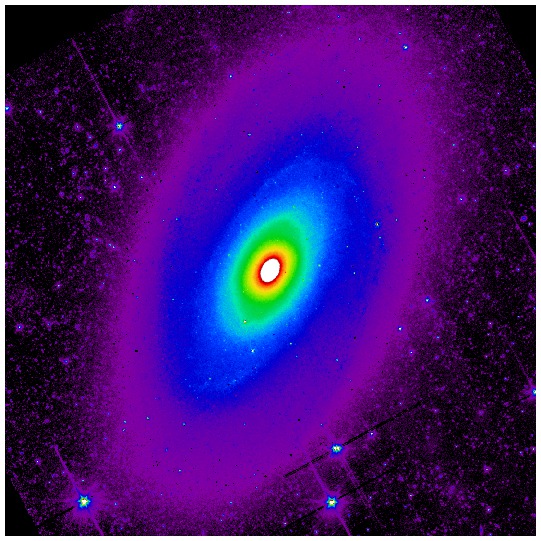}& \includegraphics[width=1\linewidth]{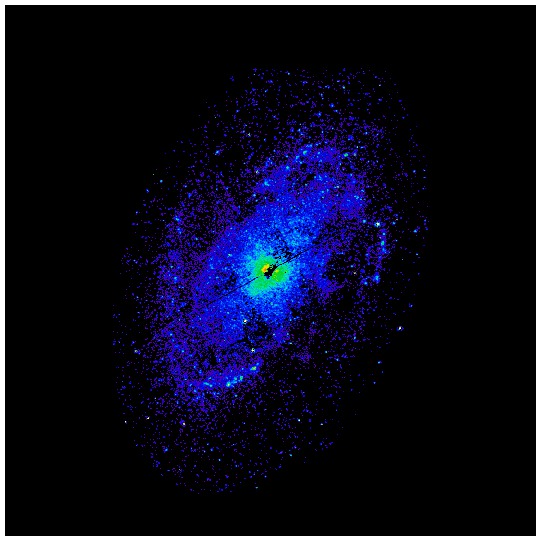}&
\includegraphics[width=1\linewidth]{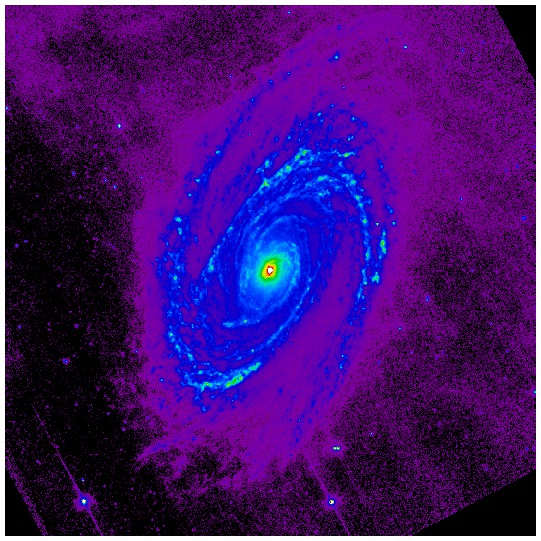}\\
\rotatebox{90}{\footnotesize NGC 3184}& \includegraphics[width=1\linewidth]{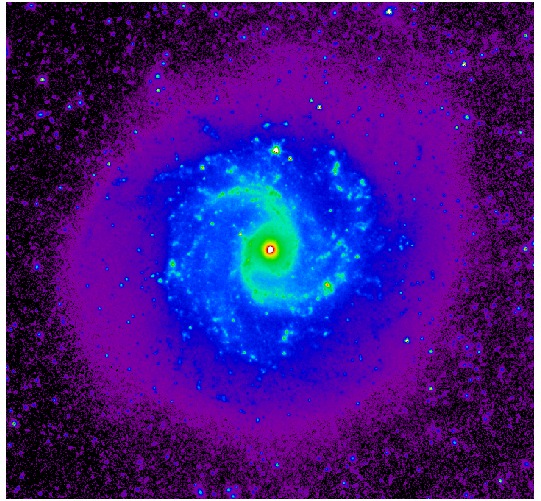}& \includegraphics[width=1\linewidth]{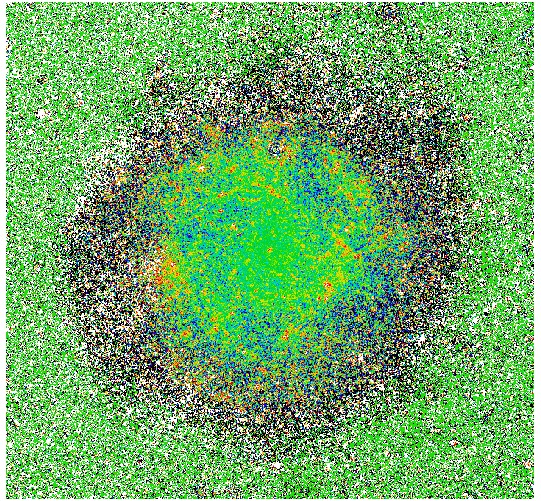}&
\includegraphics[width=1\linewidth]{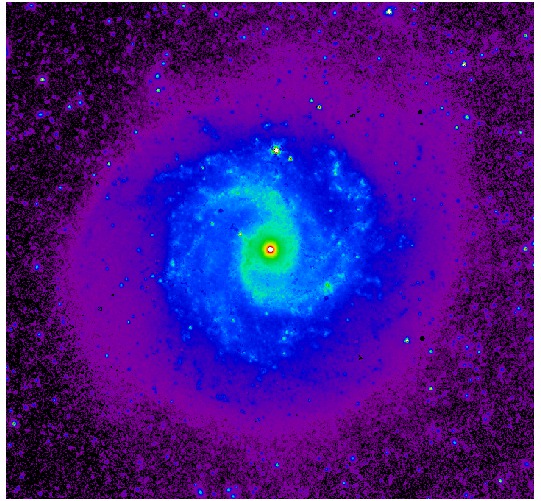}& \includegraphics[width=1\linewidth]{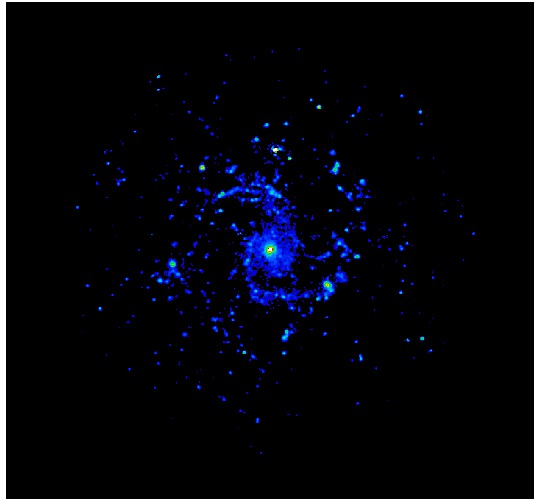}&
\includegraphics[width=1\linewidth]{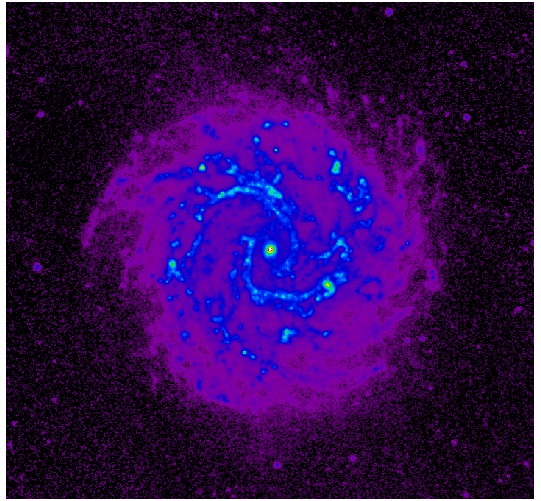}\\
\rotatebox{90}{\footnotesize NGC 4321}&  \includegraphics[width=1\linewidth]{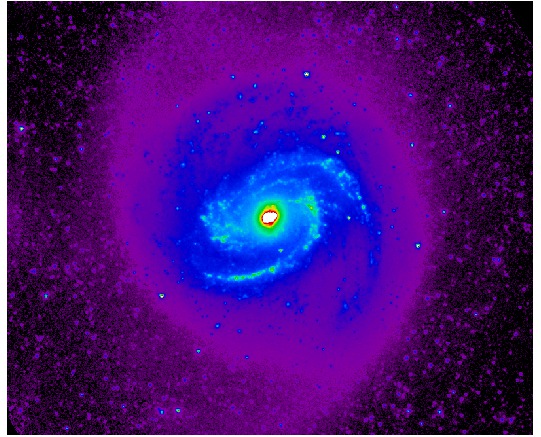}& \includegraphics[width=1\linewidth]{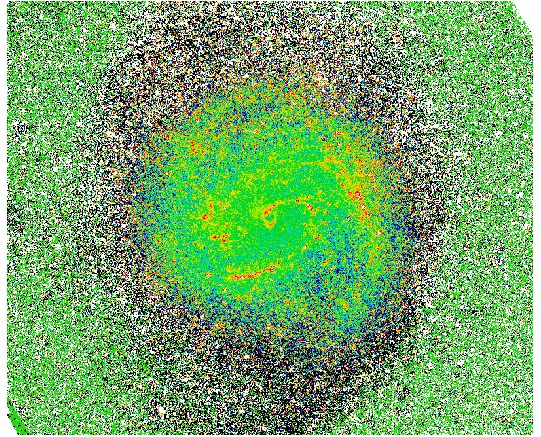}&
\includegraphics[width=1\linewidth]{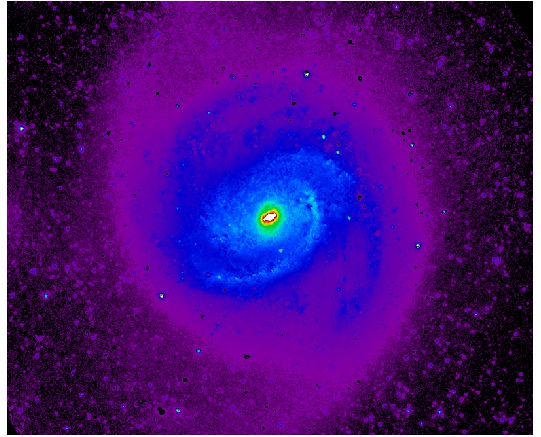}& \includegraphics[width=1\linewidth]{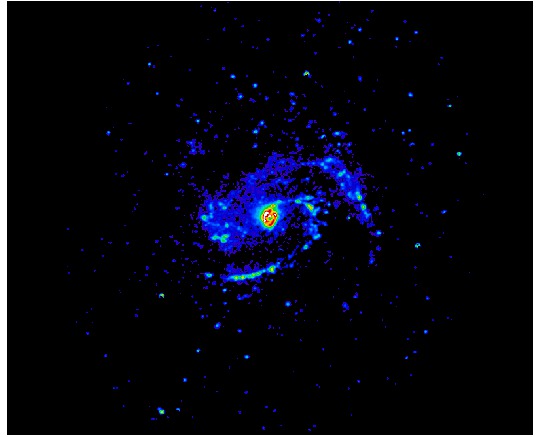}&
\includegraphics[width=1\linewidth]{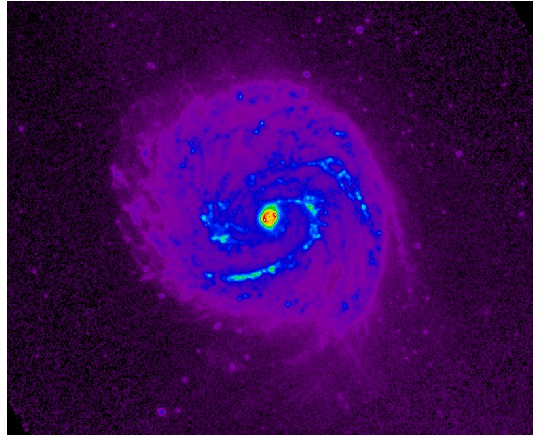}\\
\rotatebox{90}{\footnotesize NGC 5194}& \includegraphics[width=1\linewidth]{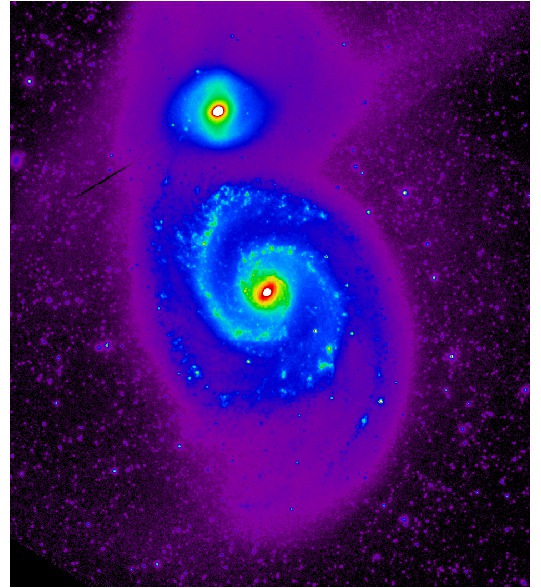}& \includegraphics[width=1\linewidth]{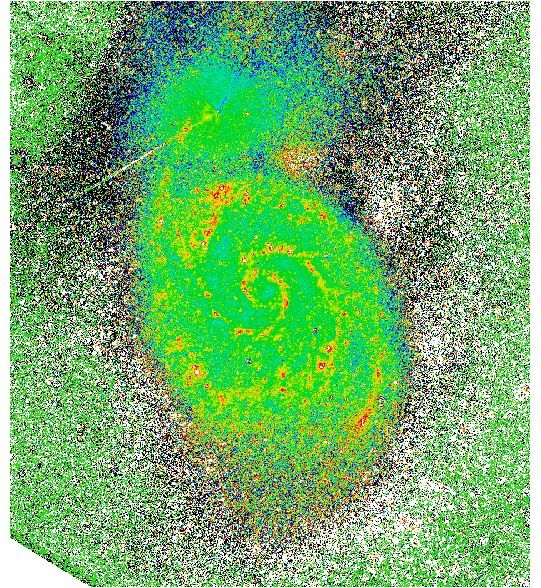}&
\includegraphics[width=1\linewidth]{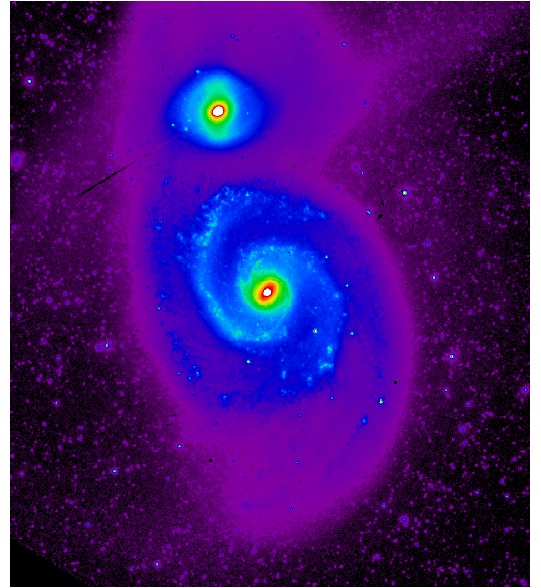}& \includegraphics[width=1\linewidth]{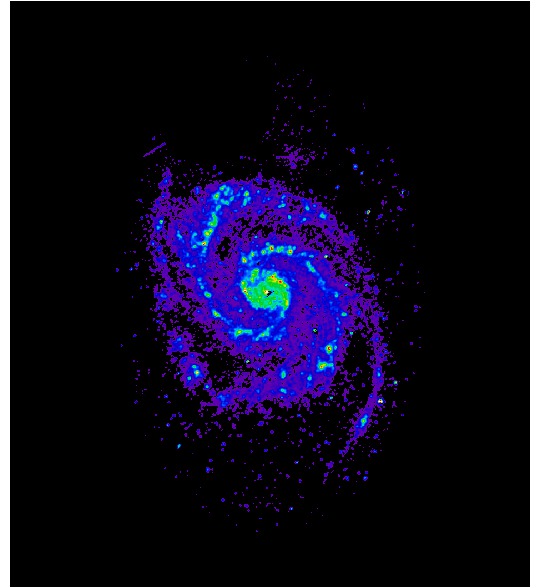}&
\includegraphics[width=1\linewidth]{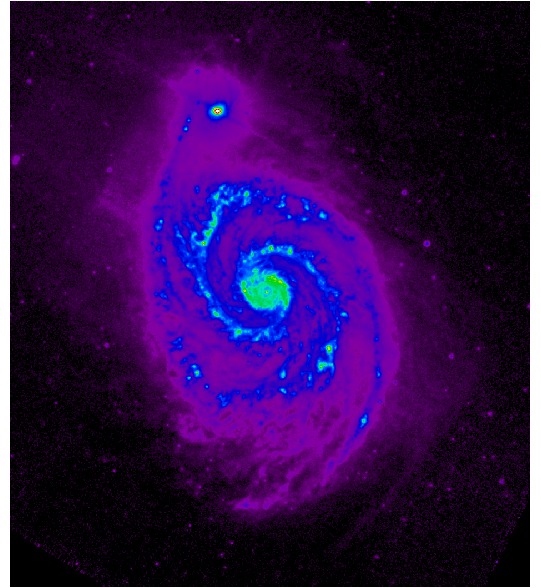}\\
\end{tabular}
\caption{\small From left to right: S$^4$G pipeline-processed IRAC channel 1 (3.6 $\mu m$) image \citep{sheth}; [3.6]-[4.5] color map; separation of sources achieved with ICA at 3.6 $\mu m$, map s1-ICA (old stars) and map s2-ICA (contaminants); SINGS IRAC channel 4 (8 $\mu m$) image (\citealt{kennSings}) shown with the stellar continuum subtracted as in $\S$ \ref{sec:dustMeasures}.  All images are shown on a square root intensity scale (range given below), except for the [3.6]-[4.5] color (shown on a linear scale from -0.3 to 0.3 mag, with [3.6]-[4.5]=-0.3 in black, [3.6]-[4.5]=-0.15 in blue, [3.6]-[4.5]=0 in green, [3.6]-[4.5]=0.15 in orange and [3.6]-[4.5]=0.3 in white).  The upper end of the intensity range $I^{max}$ for images shown in columns 3, 4 and 5 (s1-ICA, s2-ICA, and the stellar continuum subtracted 8$\mu m$ emission, respectively) is set by the value $I_{3.6}^{max}$ adopted for the original 3.6 $\mu m$ image shown in column 1 (where $I_{3.6}^{max}$=5 MJy/sr in NGC 2976 and NGC 3184, $I_{3.6}^{max}$=12 MJy/sr in NGC 4321 and NGC 1566, $I_{3.6}^{max}$=15 MJy/sr in NGC 3031 and $I_{3.6}^{max}$=20 MJy/sr in NGC 5194).  For all galaxies: $I_{s1}^{max}$=$I_{3.6}^{max}$, $I_{s2}^{max}$=0.5$I_{3.6}^{max}$ and $I_{8}^{max}$=10$I_{3.6}^{max}$, except for NGC 3031 where $I_{s2,3.6}^{max}$=0.2$I_{3.6}^{max}$=0.1$I_{8}^{max}$.  (In NGC 3031, where $L_{IR}/L_{opt}$ in Table \ref{tab-params} is lowest, contaminant fluxes are fainter, and the implied 3.6/8 $\mu m$ PAH ratio lower, than in all other galaxies; see Figure \ref{fig-scat}.)
Field dimensions (top to bottom): 10.6'$\times$8.75' (NGC 4321);  12.26'$\times$13.81' (NGC 5194); 9.51'$\times$8.96' (NGC 3184); 7.54'$\times$6.76' (NGC 2976); 10.52'$\times$8.82' (NGC 1566); 22.52'$\times$22.52' (NGC 3031).  \label{fig-4321}}
\end{flushleft}
\end{figure*}

Because the distributions in the source maps at 3.6 $\mu m$ and at 4.5 $\mu m$ are identical to within a difference in scaling (denoted by the color), we present only solutions at 3.6 $\mu m$ here.   These are shown for each galaxy in the third and fourth columns of Figure \ref{fig-4321}, hereafter s1-ICA (the old stellar component) and s2-ICA (the contaminants).  The original 3.6 $\mu m$ image is displayed together with the original map of [3.6]-[4.5] colors in first and second columns, respectively, and the 8 $\mu m$ SINGS image is shown in the last column.

In general, map s1 is a clean, smoothed version of the uncorrected image, while map s2 contains most, if not all, of the bright and knotty features tracing, e.g., the spiral morphology.   The separation has imperfections, though, such as an over-subtraction of the contaminants in isolated zones that, in some cases, leaves a `hole' in the old stellar light.  A discussion of image imperfections will be postponed until $\S$ \ref{sec:effects}.

Comparison with the [3.6]-[4.5] color maps in Figure \ref{fig-4321} demonstrates that this separation of the input distribution into two distinct emission sources is effectively a separation by intrinsic color; regions with [3.6]-[4.5]$\gtrsim$0 appear exclusively in the s2-ICA map, whereas map s1-ICA is composed of sources with [3.6]-[4.5]$\lesssim$0.  The final colors of the two sources listed in Table \ref{tab-colors} are in fact consistent with either `red' contaminant (dust) emission or `blue' stellar light dominated by K or M type giant stars (e.g. \citealt{pahre2}; \citealt{pahre}).\footnote{Old K and M giants exhibit blue colors in the IRAC bands due to photospheric CO absorption in the 4.5 $\mu m$ band.  M stars have colors [3.6]-[4.5]$\sim$-0.15, while younger K stars with weaker CO absorption appear redder ([3.6]-[4.5]$\sim$-0.06; e.g. as noted by \citealt{willner} and \citealt{pahre}).}  

These two ICA source colors are uncorrelated, as we would expect for two physically distinct emission processes.  But, even for the small sample here, each can be related to other observable disk properties.  A larger sample will be essential for investigating dependencies on dust properties, metallicity, and star formation history, but we note that stellar disks assigned the bluest [3.6]-[4.5] colors (marked by more negative [3.6]-[4.5] colors in Table \ref{tab-colors}) are those with the reddest optical colors (e.g. $g-i\sim$0.85, using the asymptotic magnitudes measured by \citealt{munozmateos}) and earliest Hubble type in the sample.  This confirms that the ICA-measured [3.6]-[4.5] colors of stellar disks reproduce the sequence noted by \citet{pahre2}, namely that stellar disks with blue [3.6]-[4.5] colors are older than their redder counterparts (in light of the modest age difference between K and M giants dominating the observed light). 

In addition, the reddening of contaminants over the range in [3.6]-[4.5] color observed here coincides with trends in the measured properties of the emitting dust in these galaxies, as compiled by \citet{draine07} using information from throughout the dust SED.  
For example, galaxies with the reddest dust (i.e. largest positive [3.6]-[4.5] in Table \ref{tab-colors}), which have lower local contributions from contaminant emission (as measured in $\S$ \ref{sec:rsgagb}), tend to have smaller dust masses and a lower fraction of dust in the form of PAHs (measured by $q_{PAH}$; see \citealt{draine07}, Tables 4 and 5).  A larger sample will be necessary to link these trends.  In what follows we first undertake a detailed study of the two ICA sources at 3.6 $\mu m$ using the spatial and intensity information in the ICA source maps. 
\begin{table}
\begin{center}
\caption{ICA [3.6]-[4.5] colors\label{tab-colors}}
\begin{tabular}{rcccc}
\tableline\tableline

 Galaxy&s1&$\delta_{s1}$&s2&$\delta_{s2}$\\
\tableline
NGC 1566&-0.132&0.021&0.669&0.242\\
NGC 2976&-0.143&0.022&0.44&0.029\\
NGC 3031&-0.102&0.015&0.239&0.194\\
NGC 3184&-0.127&0.051&0.764&0.311\\
NGC 4321&-0.071&0.019&0.242&0.062\\
NGC 5194&-0.094&0.029&0.542&0.231\\
\tableline
\end{tabular}
\tablecomments{Global [3.6]-[4.5] colors for the two ICA separated sources, s1-ICA (old stars) and s2-ICA (contaminants).   Variation in the color of each component over a fixed range of initial seeds (see text) defines the uncertainty measure $\delta$.  The larger uncertainties for source s2 reflect the larger intrinsic spread in contaminant sources collected together in the single contaminant map, as inventoried in $\S$ \ref{sec:contams}. For reference, all colors would be lower by -0.012 using the zeropoints adopted by \citet{pahre2}.}
\end{center}
\end{table}
\section{The Contaminants}
\label{sec:contams}
\indent In this section we explore how well the expected sources of contamination in the 3.6 $\mu m$ images are detected with ICA.  These contaminants are defined in what follows as all sources of emission apart from the oldest 
stars: PAH, hot dust, and intermediate-age red supergiant (RSG) and asymptotic giant branch (AGB) stars.  With the removal of these contaminants from S$^4$G images, the remaining old stellar light should be readily convertable into mass maps.  
\subsection{Emission from Dust\label{sec:dust}}
\indent The apparent similarity between maps s2 and the 8 $\mu m$ images for all galaxies is a first strong indication that the emission in the ICA-calculated map can be associated with non-stellar emission.   A more strict identification of, e.g., the expected 3.3 $\mu m$ PAH contaminant there is complicated, however, given that the scaling between this feature and the standard 8 $\mu m$ PAH proxy depends on the size and ionization state of the PAH \citep{bakes}, and is thus fairly uncertain.  

Emission from hot dust as traced, e.g., by 24$\mu m$ images, which is present to varying degrees in these galaxies (see the 8/24 $\mu m$ flux ratio in Table \ref{tab-params}), also contributes at 3.6 $\mu m$ and 4.5 $\mu m$.  Hot dust emission is well-correlated with 8 $\mu m$ PAH emission only on the largest scales.  Below a few hundreds of parsecs (on the scale of HII regions) \citet{bendo} find an excess of hot dust emission relative to the PAH. 

These two factors are of critical importance for interpreting the s2 contaminant map.  
Emission from both hot dust and PAH (the 3.3 $\mu m$ feature in the 3.6 $\mu m$ band and continuum at 4.5 $\mu m$; i.e. \citealt{flagey}) contaminate the 3.6 and 4.5 $\mu m$ images whereas, in performing ICA with only 2 bands, we assume only a single non-stellar source.  This effectively supposes that the PAH emission scales perfectly with the hot dust, i.e. given a global scaling between the 3.3 $\mu m$ PAH emission (and continuum) and the broad component at 8 $\mu m$, together with an equivalently uniform scaling between the 8 $\mu m$ emission and the hot dust as imaged at 24$\mu m$.   

In this case, only the dust spectral index given by the ratio of the s2 contaminant map at 3.6 and at 4.5 $\mu m$ ($L_{s2,3.6}$/$L_{s2,4.5}$) would deviate from what we expect based on thermal dust emission alone.  We find this to be the case in all our maps.  The colors for source s2 listed in Table \ref{tab-colors} are bluer than the [3.6]-[4.5]$\sim$0.97 color characteristic of power-law emission on the Wien side of the dust SED, with spectral index $n_{hot dust}$=-2 \citep{blain}.  The PAH-to-hot dust ratio is evidently close enough to globally constant that it can be modeled as such, even if there may be real local deviations (i.e. in HII regions).  Letting ICA choose the optimal balance, these deviations remain recorded within our uncertainty measure $\delta_{s2}$, which serves to quantify the risk we inherit in the limit of detecting only one contaminant where more than one source of emission may exist (see e.g. $\S$ \ref{sec:rsgagb}).

This simple scenario also holds well enough that we find a good correlation between the contaminant flux at 3.6 $\mu m$ and the 8 $\mu m$ emission, as demonstrated below.   Even when galaxies exhibit extreme spatial variations in the PAH-to-hot dust ratio, and the `perfect scaling' scenario becomes less appropriate, the effect on both the ICA-identified contaminants and old stars is small, within the (1$\sigma_m$) photometric errors.  
\subsubsection{Comparison with 8 $\mu m$ PAH emission\label{sec:dustMeasures}}
\indent Below we explore the correlation between the emission in map s2 and emission from dust as imaged at 8 $\mu m$, which is dominated by the broad PAH feature.  While we do not expect a simple, one-to-one relation (given the factors discussed above, as well as the inevitable uncertainty in the subtraction of the stellar component from the 8 $\mu m$ map), the trends are useful for distinguishing between hot dust, PAH emission and `red knots' dominated by intermediate-age stars (e.g. $\S$ \ref{sec:rsgagb}).  In addition, this exercise may motivate future initiatives to study PAH and hot dust emission within the s2 contaminant maps; these maps of the dust have higher angular resolution (i.e. 1.2'' in 3.6 and 4.5 $\mu m$ images) than maps at longer wavelengths. 

Finally, comparison to the 8$\mu m$ emission provides as accurate quantification of the quality of our contaminant correction as is currently possible, given that corrections for the contaminants in the IRAC bands are typically based on such images, where available (\citealt{kendall}; \citealt{helou}).   Because we prefer to retain the high native resolution at 3.6 $\mu m$, the pixel-by-pixel comparisons to the 8 $\mu m$ emission are not performed on PSF-matched images.  We have confirmed that the general conclusions drawn here are unchanged when the 3.6 and 8 $\mu m$ images are at the same resolution (the PSFs at 3.6 and 8 $\mu m$ are similarly sized, differing by a factor of $\sim$1.2 in FHWM), suggesting that it is not necessary to degrade the 3.6 $\mu m$ resolution to make identifications similar to those described in this section.   Later in $\S$ \ref{sec:rsgagb}, though, we comment on circumstances beyond our immediate concern that require a careful treatment of PSF differences.  \\
\begin{figure*}[htbp] 
   \centering
 \plottwo{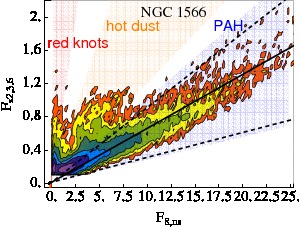}{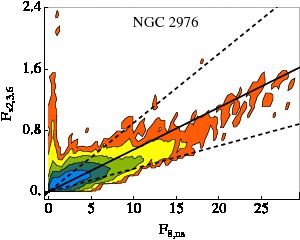}   
 \plottwo{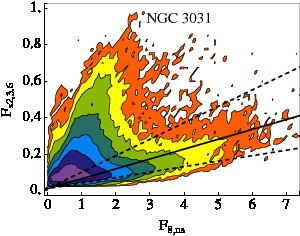}{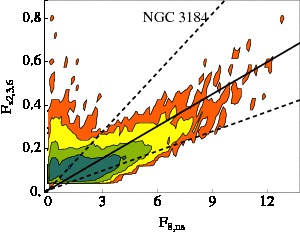}
   \plottwo{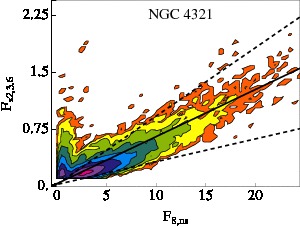}{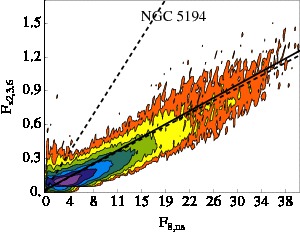}
      \caption{\small Contour plots of $F_{s2,3.6}$ vs. $F_{8,ns}$ (calculated as described in the text) for NGC 1566 (top left), NGC 2976 (top right), NGC 3031 (middle left), NGC 3184 (middle right), NGC 4321 (bottom left) and NGC 5194 (bottom right), with number of image pixels increasing from the outermost (orange) contour to the innermost (purple) contour.  Over-layed dashed lines represent the expected range in $F_{s2,3.6}$=$\alpha$$F_{8,ns}$ for PAH emission, with $\alpha$=0.09 or 0.03.  A third solid line shows the relation given by the best-fit slope $\hat{\alpha}$ listed in Table \ref{tab-pah}.   \label{fig-scat}}
\end{figure*}
In Figure \ref{fig-scat} we plot for each galaxy the flux measured in s2 against the non-stellar flux at 8 $\mu m$ where, for all galaxies, the non-stellar flux at 8 $\mu m$ is calculated as $F_{8,ns}=F_{8}-0.232 F_{s1,3.6}$ based on the stellar continuum correction given by \citet{helou}, and with the convention $F$=$f_{\nu}$ (also used for all remaining fluxes).  Note that in this definition we represent the stellar component with our map s1 at 3.6 $\mu m$.   

Most galaxies (with the exception of NGC 3031 and NGC 1566; see $\S$ \ref{sec:hotdust}) show a clear relation between the flux in s2 and the PAH emission at 8 $\mu m$ over almost the full range in $F_{8,ns}$.  Notably, the slope of the implied linear relation for each falls within the range of expected values for the scaling $\alpha$ between the PAH emission at 3.3$\mu m$ and at 8 $\mu m$, 0.03$\lesssim$$\alpha$$\lesssim$0.09, as measured in the Milky Way (see \citealt{flagey} and references therein).  At the very least, this suggests that the ICA-identified component is a reasonable map of the PAH emission at 3.6 $\mu m$, by virtue of the realistic fluxes assigned to s2 (i.e. the ICA `measured' fluxes are consistent with a dust component).   There is obvious scatter, though, and the overall relation in some cases features a shifting slope with increasing or decreasing $F_{8,ns}$.  

The scatter in the correlation might be partly attributed to differences in the size and ionization state of the PAHs throughout the disk.  But in addition to physical variations in the PAH, uncertainty in the stellar continuum subtraction (i.e. given error in s1, or in the scaling of the old stellar light from 3.6 to 8 $\mu m$), a non-dust origin for some of the `contaminant' flux in s2, or thermal dust emission could plausibly contribute to the overall scatter in the panels of Figure \ref{fig-scat}.  The first case does not seem to be a major source for this small sample; we find no significant change in the scatter or overall appearance of these plots with $F_{s1,3.6}$ replaced by the original, uncorrected fluxes in the definition of $F_{8,ns}$ above.  In the second case, the signature of a non-dust source of contamination is more obviously manifest to the left in the plots, where $F_{s2,3.6}$ is in excess above $F_{8,ns}$, as considered further in the section $\S$ \ref{sec:rsgagb}.  

The more likely contributor to the plots in Figure \ref{fig-scat} where $F_{8,ns}$$\gtrsim$2 MJy/sr (i.e. any trend for a variable slope, but also some scatter), and especially to the overall appearance in the case of NGC 3031, would seem to be the non-PAH (hot dust) component in both the 8 $\mu m$ and 3.6 $\mu m$ bands, as demonstrated in the following section.  

Another possible source of correlation between $F_{s2,3.6}$ and $F_{8,ns}$ could arise with our definition of $F_{8,ns}$, although we do not find strong evidence for such a possibility here.  Specifically, using $F_{s1,3.6}$ rather than $F_{3.6}$ to subtract the stellar continuum contribution in $F_8$ introduces in $F_{8,ns}$ an explicit dependence on $F_{s2,3.6}$ (i.e. $F_{8,ns}$=$F_{8}$-0.232$F_{3.6}$+0.232$F_{s2,3.6}$), potentially leading to a forced correlation at fixed observed $F_8$ and $F_{3.6}$.   However, the latter two observables are far from fixed, and the variance in the quantity $F_{8}$-0.232$F_{3.6}$ is at least an order of magnitude larger than the variance in $F_{s2,3.6}$ in all cases.  (In one example, NGC 4321, the variance in $F_{8}$-0.232$F_{3.6}$ is 13.6 MJy/sr, while the variance in $F_{s2,3.6}$ is 0.6 MJy/sr.)  Meanwhile, the variance in the constructed $F_{8,ns}$ is comparable to the variance in $F_{8}$-0.232$F_{3.6}$ (e.g. in NGC 4321 the variance in  $F_{8,ns}$ is 13.4 MJy/sr), which strongly indicates that $F_{8,ns}$ is driven primarily by $F_8$ and $F_{3.6}$, not $F_{s2,3.6}$.  Likewise, we see very little change in the appearance of these plots with $F_{s1,3.6}$ replaced by $F_{3.6}$ in the definition of $F_{8,ns}$, as mentioned earlier.   Altogether, this supports the conclusion that $F_{8,ns}$ depends only negligibly on $F_{s2,3.6}$, or that the correlations between $F_{s2,3.6}$ and the imperfectly independent variable $F_{8,ns}$ are not the result of the (minor) dependence of $F_{8,ns}$ on $F_{s2,3.6}$.  Instead, these correlations arguably arise from the physical characteristics of the emission contained in map s2.
\subsubsection{Contribution from hot dust\label{sec:hotdust}}
Comparison with the non-stellar emission at 8$\mu m$ demonstrates not only the detection of PAH emission in the contaminant maps at 3.6 $\mu m$ but also a second, hot dust contaminant.
Consider, for instance, the grouping of pixels at low 8 $\mu m$ flux ($F_{8}$$\lesssim$3 MJy sr$^{-1}$) in either NGC 1566 or NGC 3031 following a steepened trend.   In both cases, these pixels can be identified with regions in the disk where the ratio $F_{8}$/$F_{24}$ drops, and hot dust dominates the emission from PAHs.  In NGC 1566, for example, these pixels cluster in small isolated HII regions that string along the spiral arms, down in to the center of the disk (see Figure \ref{fig-contams}).  In NGC 3031, these pixels are located in the central 75'', where $F_{8}$/$F_{24}$ is low in the surrounds of NGC 3031's AGN (see Figure \ref{fig-contams}; e.g. \citealt{bendo}).  In the immediate vicinity of the unobscured nucleus in this particular case, the mid-IR emission can be associated with the signature ``big red bump" emission from a truncated accretion disk (\citealt{ho}; \citealt{horev}; \citealt{quataert}), but emisson from an extended hot dust component as far out as 75'' (possibly heated by the AGN) presumably also contributes to the drop in $F_{8}$/$F_{24}$ here.  

In these two cases we expect that the steepened correlations in Figure \ref{fig-scat} are dictated less by the PAH than by the other main source of mid-IR emission, given the observed low level non-stellar $F_{8}$/$F_{24}$.   In this case, the ratio $F_{s2,3.6}$/$F_{8,ns}$ at each pixel should roughly measure the non-stellar spectral index $n$ (but see the upcoming discussion).  Using the measured ratio $F_{s2,3.6}$/$F_{8,ns}$ we find a spectral index consistent with $n$=2 (\citealt{blain}; as expected for power-law emission on the Wien side of the dust spectrum) with which we recover the observed flux at 24$\mu m$ within 15\%, on average, in both NGC 1566 and NGC 3031.   

A large spread in the flux of hot dust over the disk will result in a `continuum' of shifted linear relations between $F_{s2}$ and $F_{8,ns}$ for which $\alpha$ may also vary, possibly contributing to the overall scatter in these plots.  But because the fall-off rate of hot dust emission between 8 and 3.6 $\mu m$ is smaller than ratio of the PAH features in these bands, the signature of the hot dust will in general appear systematically to the left of the PAH in the $F_{3.6}$-$F_{8}$ space plotted in Figure \ref{fig-scat}, making this component straightforward to identify in the contaminant map.   As depicted in Figure \ref{fig-contams}, the range 0.1$\lesssim$$F_{s2,3.6}$/$F_{8}$$\lesssim$0.3 can be used to identify pixels with emission from hot dust, while the criterion $F_{s2,3.6}$$\lesssim$0.09 selects for PAH emission.  

The spatial locations of hot-dust dominated regions can be successfully identified in this way.  Hot dust fluxes may be slightly less reliable without correction, though, as they are susceptible to overestimation.  
Since the spatially predominant PAH component often more heavily controls the contaminant color than the hot dust, then the ICA [3.6]-[4.5] color is slightly too blue in areas with the highest contribution from this source (i.e. in HII regions).\footnote{Whereas hot dust should have [3.6]-[4.5]$\sim$0.9, for the PAH we expect [3.6]-[4.5]$\sim$0.04-0.2 according to the measured PAH contributions in the 3.6 and 4.5 bands (see, e.g., \citet{flagey}) and as is typical for the [3.6]-[4.5] color maps in Figure \ref{fig-4321}.}  
This leads to an overestimation in the flux from the hot dust there, which would nominally appear more red in the [3.6]-[4.5] color.   But as long as the flux in hot dust  $F_{3.6}^{HD}$ is much larger than the flux from PAHs, the fractional overestimation in $F_{3.6}^{HD}$ will be independent of $F_{3.6}^{HD}$ and proportional to the color difference between the ICA value and the true value, leaving a well-defined trend in 3.6-8 $\mu m$ space identifiable with hot dust emission.  

Conversely, as discussed more thoroughly in $\S$ \ref{sec:imperfections}, the presence of hot dust can contribute to the underestimation of the PAH emission throughout the rest of the disk, but only most strongly in cases where the ICA [3.6]-[4.5] color nears that of the hot dust.  The color of the intrinsically bluer PAH emission would then be over-predicted, resulting in an underestimation of the flux from this component.  

Even in cases where the hot dust becomes more dominant overall here we find little evidence that the PAH is affected, and the picture of the dust presented by the contaminant maps remains physically realistic.  Galaxies with the reddest contaminant color are not those with the lowest $\alpha$, for instance, although we note that the coincidence of a low overall $\alpha$ with evidence for a second (e.g. hot dust) contribution, for instance, would still be arguably consistent with a link between the ionization state of the PAH (as indicated by $\alpha$; see \citealt{bakes}) and the amount of 24$\mu m$ emission, with dust temperature the commonality. Plus, at least in this small sample, the ICA contaminant flux is arguably the fairest balance between the PAH and the hot dust in these maps where the latter component is most prominent.  Emission in NGC 3031 is divided into almost equal subsets of PAH and hot dust, for example, whereas the majority of emission in NGC 4321 is consistent with PAH emission (see Figure \ref{fig-scat}).   

Based on the demonstrated correlation with the 8 $\mu m$ emission, we argue that map s2 provides a realistic description for the state of the dust in the studied galaxies.  
Uncertainties in the quality of the resulting correction to 3.6 $\mu m$ images should be kept in mind, however, given the modest drawback to detecting the dust features together as noted earlier in this section (and addressed again in $\S$ \ref{sec:multidists}).  
\begin{table}
\begin{center}
\caption{PAH Contamination between 3.6 and 8 $\mu m$\label{tab-pah}}
\begin{tabular}{rcl}
\tableline\tableline

Galaxy&$\hat{\alpha}$&fit zone\\
\tableline
NGC 1566&0.065$\pm$0.0001&$F_{8,ns}$$>$5 MJy/sr\\
NGC 2976&0.068$\pm$0.0002&$F_{8,ns}$$>$1.5 MJy/sr\\
NGC 3031&0.036$\pm$0.0001&$L_{8,ns}$$>$2.5 MJy/sr\\
NGC 3184&0.049$\pm$0.0001&$F_{8,ns}$$>$2.5 MJy/sr\\
NGC 4321&0.062$\pm$0.0002&$F_{8,ns}$$>$2 MJy/sr\\
NGC 5194&0.031$\pm$0.0002&$r$$<$$225''$, $F_{8,ns}$$>$3 MJy/sr\\
\tableline
\end{tabular}
\tablecomments{Parameters of the best fit straight line $F_{s2,3.6}$=$\hat{\alpha}F_{8,ns}$ in the scatter plots shown in Figure \ref{fig-scat}, with selection criteria for the fit regions listed on the right.  These fitted regions are chosen by eye to avoid emission from hot dust and red objects (see $\S$ \ref{sec:rsgagb}) in order to most fairly portray the behavior of the PAH emission between 3.6 and 8 $\mu m$, expected to have $\alpha$=$F_{s2,3.6}$/$F_{8,ns}$ in the range 0.03$<$$\alpha$$<$0.09.
}
\end{center}
\end{table}
\subsection{RSG and AGB stars\label{sec:rsgagb}}
\begin{figure*}[htbp] 
   \centering
\epsscale{.9}
\plottwo{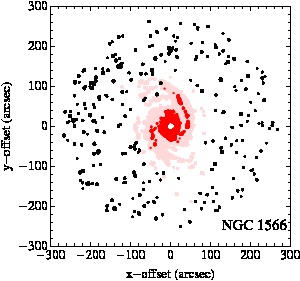}{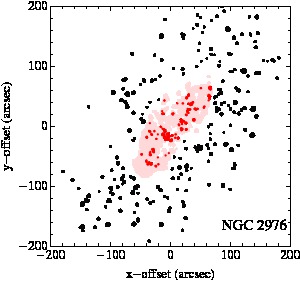} 
\plottwo{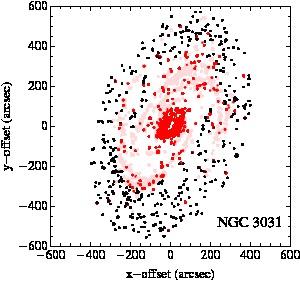}{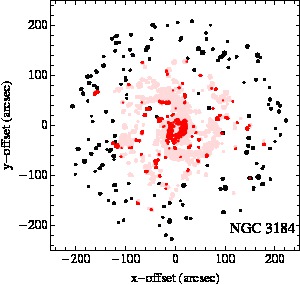}
\plottwo{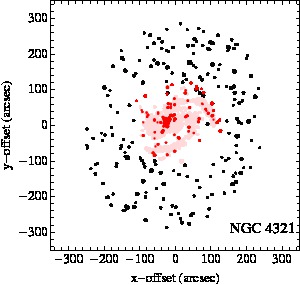}{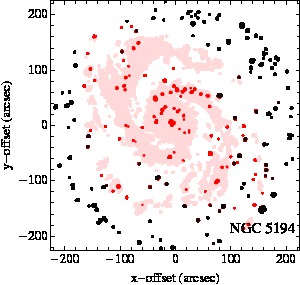} 
   \caption{\small Map of pixels representative of the `dust' and `red knot' components in the ICA contaminant distributions for NGC 1566 (top left), NGC 2976 (top right), NGC 3031 (middle left), NGC 3184 (middle right), NGC 4321 (bottom left) and NGC 5194 (bottom right; only the inner $r$$<$$225''$ is shown).  Emission from PAH (light gray), hot dust (dark gray) and 'red knots' (black) is selected from the range $F_{s2,3.6}$/$F_{8,ns}$$<$0.09,  0.1$<$$F_{s2,3.6}$/$F_{8,ns}$$<$0.3, and $F_{s2,3.6}$/$F_{8,ns}$$>$0.3, respecitvely.   \label{fig-contams}}
 \end{figure*}
\begin{figure*}[htbp] 
   \centering
\epsscale{.9}
\plottwo{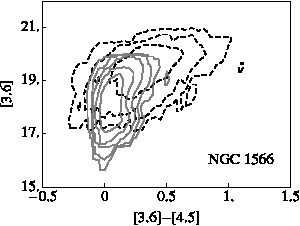}{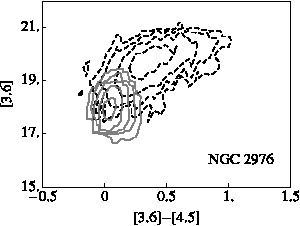} 
\plottwo{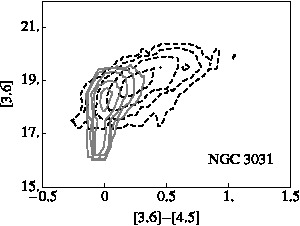}{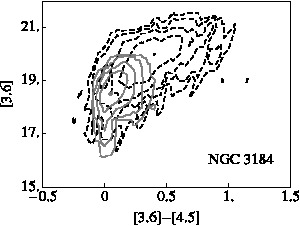}
\plottwo{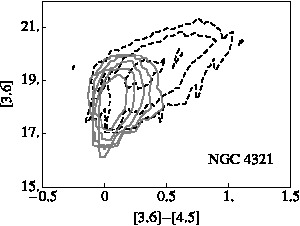}{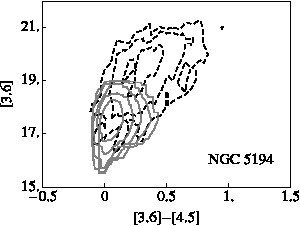}
   \caption{\small CMD for pixels representative of the primary `dust' ($F_{s2,3.6}$$<$0.3$F_{8,ns}$; in gray solid line, including emission from both PAH and hot dust) and secondary `red knot' ($F_{s2,3.6}$$>$0.3$F_{8,ns}$; in black dashed line) components in the ICA contaminant distributions for NGC 1566 (top left), NGC 2976 (top right), NGC 3031 (middle left), NGC 3184 (middle right), NGC 4321 (bottom left) and NGC 5194 (bottom right). (See text for selection criteria.)   \label{fig-cmds}}  
\end{figure*}

Next we reveal an additional source of contamination in ICA map s2 besides the combination of hot dust and the 3.3 $\mu m$ PAH emission, namely clusters that host bright but low M/L evolved AGB and RSG stars contributing to the emission at this wavelength.  
Such evolved objects are typically clearly identifiable by their location in [3.6] vs. [3.6]-[4.5] color-magnitude diagrams (CMDs) (c.f. \citealt{jackson}; \citealt{davidge}) given their characteristically (though not necessarily uniform) `red' [3.6]-[4.5]$>$0 colors.\footnote{\indent Dusty, mass-losing RSGs will have colors in excess of the blue [3.6]-[4.5] color exhibited by other stars of K and/or M spectral type, depending on metallicity $Z$ (mass loss and dust production increase with increasing $Z$).   Even the youngest of these may then appear within the contaminant map where the effect of the dust in the IRAC bands exceeds the CO absorption at 4.5 $\mu m$ (which also increases with increasing $Z$).  (See \citet{bonanos}, \citet{jackson} and references therein.)  }  
In the ICA-corrected maps, on the other hand, colors are at a constant value [3.6]-[4.5]=-0.1$\pm$0.03 on average (see Table \ref{tab-colors}), comparable to what is expected for an M giant \citep{pahre2}.   By extension, if regions dominated by evolved RSG and AGB stars are present in the original 3.6 $\mu m$ image, they can be identified with ICA as part of the `contaminant' distribution in the s2 image. 

Qualitative comparisons between the contaminant map and the 8 $\mu m$ image are often alone suggestive of a distinct secondary component.  This is especially obvious for NGC 3031, where the ICA map s2 exhibits flux in locations distinctly apart from the (strongly) emitting zones in the 8 $\mu m$ image.  
Below we introduce a more quantitative characterization of the second component in the ICA contaminant distribution by combining the results of the previous section with spatial and color information. 

To start, pixels originating in regions with little to no 8 $\mu m$ dust emission can be easily identified in the plots of Figure \ref{fig-scat} as a thin, almost vertical concentration of points at the far left ($F_{8,ns}$$\lesssim$2 MJy/sr).   These can be isolated using a coarse selection $F_{s2,3.6}$$\gtrsim$0.3$F_{8}$, avoiding the zone in 3.6-8$\mu m$ space populated by hot dust near $F_{s2,3.6}$/$F_{8}$$\sim$0.2 (for power-law emission on the Wien side of the dust spectrum with index $n$=2, following \citealt{blain}, as discussed in the previous section).  
The locations of a representative set of these pixels (i.e. only those above $F_{s2,3.6}$$\sim$0.2 MJy/sr, for the purposes of clarity) for all galaxies are shown in black in Figure \ref{fig-contams}.  To minimize contamination by obvious foreground stars, the selection includes an upper limit on the flux at the position of candidate pixels in the original image.  

The spatial configuration of the black pixels in these maps is clearly distinct from the dust, represented in light gray (PAH) and dark gray (hot dust), but in some cases similar to what we might expect for a random distribution of field stars and background galaxies.  (Recall that no masking of such objects has been invoked.)  While external contaminants should in large have colors [3.6]-[4.5]$\sim$0, we find mostly colors 0$\lesssim$$[3.6]$-$[4.5]$$\lesssim$1.0 for the second distribution of contaminant pixels.

This is clear in Figure \ref{fig-cmds} showing representative color-magnitude diagrams for the dust and `red knot' components in s2--separated as described above--which we construct using the original color and magnitude at each pixel position in the uncorrected images (see the second column in Figure \ref{fig-4321}).  
In each plot, the two distributions populate different, yet overlapping regions in this color-magnitude space:  
pixels with $F_{s2,3.6}$$>$0.3$F_{8}$ extend to fainter magnitudes and have redder [3.6]-[4.5] colors (at a given magnitude) than pixels with $F_{s2,3.6}$$<$0.3$F_{8}$.  

According to typical CMD identifications (\citealt{jackson}; \citealt{davidge}) the red colors of the former set 
are consistent with evolved, post-MS stars  (and with predictions for AGB stars, in particular, the most luminous with 0.5$<$[3.6]-[4.5]$<$1; \citealt{groenewegen}).   These regions could also be sites of very recently formed massive star clusters prevented from dispersal that illuminate the surrounding diffuse dust (e.g. \citealt{hunter}) and so appear red in the IRAC bands.  A subset are also likely identifiable with nearby background galaxies for which the observed mid-IR colors are closest to rest-frame colors (e.g. \citealt{fazio}). These possibilities will be considered in greater detail in future work using archival broadband data to construct the optical-to-mid-IR SEDs in these regions (Meidt et al., in prep).  Here we note that preliminary analysis in NGC 4321 suggests these objects are intrinsically red, with optical colors consistent with AGB stars (\citealt{jackson}; V-I $\sim$1.5). In addition, the [3.6]-[8] colors in these regions are indicative of emission from circumstellar dust, as expected around AGB stars undergoing high mass loss \citep{jackson}.  

In any case, it is clear that this second contaminant population is distinct from the greater bulk of PAH and hot dust emitting at 3.6 $\mu m$ as depicted in Figure \ref{fig-contams}, and that a set of relatively young stars (with ages between 30 Myr and 1 Gyr) dominates the emission. 
Note, though, that the maps in Figure \ref{fig-contams} may strongly reflect a selection bias against spatially coincident dust and clusters dominated by intermediate-age stars.  At the extreme, the redder component in the contaminant map for NGC 2976, for instance, is found preferentially outside the bulk of the dust emission.   This likely arises with our identification of red clusters in 3.6-8 $\mu m$ space (Figure \ref{fig-scat}), which requires low values of $F_{8,ns}$ (i.e. little emission from other dust sources) to detect pixels with the highest $F_{s2,s.6}$/$F_{8,ns}$.  A more sophisticated selection would be required to map this type of pixel where, e.g., $F_{s2,s.6}$ is high and $F_{8,ns}$ is not simultaneously low.  

Despite this over-simplicity, the separation criteria developed here are suitable for obtaining rough estimates of the individual contributions of the two contaminant components to the observed emission, as listed in Table \ref{tab-contamfraction}.   Entries there indicate the fraction of contaminant emission per pixel averaged over $N$ pixels, where $N$ is the number of contaminant pixels in the case of 'local' measurements (columns 3 and 5), or the number of all pixels in the image out to the radius given in Table \ref{tab-params} in the case of 'total' (columns 2 and 4).  Note that the derived fractions are  influenced by location-dependent effects, i.e. a reduced contribution from old stellar light at larger radii.\footnote{Given the underlying radial decrease in the old stellar light, the pixel positions mapped in Figure \ref{fig-contams} must be taken into consideration when comparing the 'local' values for the two components.  (In NGC 2976, for example, the contribution of the old stellar light to pixels with emission from AGB/RSG stars is much lower than for pixels with dust emission.)}  In addition, although averaging should reduce the effect of lingering cross-contamination in the two components, an additional bias may stem from a lack of PSF-matching between the 3.6 and 8 $\mu m$ images; the 8$\mu m$ comparison directs our association of contaminant pixels with one component or the other.  Each measured fraction also inherits the risk that more than one source contaminates any given pixel.   We quantify this using the uncertainty measure for each contaminant color listed in Table \ref{tab-colors}, assuming that the color uncertainty is driven entirely by the 3.6 $\mu m$ flux; for each fraction $f$, the error $df$=$f\delta_{s2}$ by this definition.   

As expected, we find that dust can contribute between $\sim$3 to 10\% of the total emission imaged at 3.6 $\mu m$, while light from intermediate-age, post-MS stars accounts for only $\sim$1\%.  Locally, however, the contribution from either contaminant to the observed surface brightness can be much higher.  
We note that the local measurements in Table \ref{tab-contamfraction} in particular may be biased high; as commented on later in $\S$ \ref{sec:effects}, the contribution in regions with intrinsic colors much in excess of the ICA contaminant color (e.g. those dominated by AGB/RSG stars) will tend to be overestimated in the contaminant map.  The dominant old stellar light keeps this fractions accurate to within $\sim$10\%, though.   In addition, variation in the level of contamination from galaxy to galaxy may arise with the different distant-dependent physical scales being sampled, especially for the point-like contribution from intermediate-age clusters.    
A more careful evaluation of the contaminants is beyond the scope of the present work.  Currently, it is clear that 3.6 $\mu m$ emission from the PAH, hot dust, or red cluster components can be individually studied, and with the additional benefit of PSF-matching or aperture photometry for accurate flux measurement.  \begin{table*}
\begin{center}
\caption{Emission contributed by contaminants\label{tab-contamfraction}}
\begin{tabular}{rcccc}
\tableline\tableline
 Galaxy&s2, dust&&s2, stars&\\
&total&local&total&local\\
\tableline
NGC 1566&0.1$\pm$0.024&0.34$\pm$0.082&0.01$\pm$0.002&0.7$\pm$0.17\\
NGC 2976&0.17$\pm$0.005&0.28$\pm$0.008&0.01$\pm$0.0003&0.59$\pm$0.017\\
NGC 3031&0.05$\pm$0.009&0.08$\pm$0.016&0.001$\pm$0.0002&0.43$\pm$0.083\\
NGC 3184&0.03$\pm$0.009&0.19$\pm$0.059&0.007$\pm$0.002&0.24$\pm$0.075\\
NGC 4321&0.1$\pm$0.006&0.29$\pm$0.018&0.01$\pm$0.0006&0.84$\pm$0.052\\
NGC 5194, $r$$<$225''&0.08$\pm$0.018&0.21$\pm$0.048&0.001$\pm$0.0002&0.38$\pm$0.088\\
\tableline
\end{tabular}
\tablecomments{\small Average fraction of the `total' (columns 2 and 4) and `local' (columns 3 and 5) emission contributed by the two components in map s2 representing dust and intermediate-age RSG and/or AGB stars, with `total' and `local' as defined in the text.  The stellar (dust) component here is identified with the selection $F_{s2,3.6}$$>$0.3$F_{8,ns}$ ($F_{s2,3.6}$$<$0.3$F_{8,ns}$), similar to Figures \ref{fig-cmds} and \ref{fig-contams}.  Errors are defined as given in the text.  }
\end{center}
\end{table*}
\subsection{On the identification of multiple distributions\label{sec:multidists}}
In the previous two sections we clearly identified two main sources of contamination in maps s2, dust and intermediate-age, post main sequence stars.  Note the implication, in this case, that the contaminants are detected together as a single, second source.  
Although this is not necessarily guaranteed by the method, the underlying distributions of the emission components in this particular application offer some insight into such a scenario.  Consider that the fraction of pixels containing emission from the old stellar disk is much larger than the fraction of pixels containing significant emission from a contaminant.   In addition, we might expect the contaminants to be more sparsely defined within a given image, as well as possibly from image to image, than the old stars.  The combination of these two conditions would seem to make the distribution of old stars more readily definable than the distribution of contaminants, which then follows de facto.  

Only with a third measurement sample (e.g. another input image) is it possible to separate s2-ICA into two independent sources (dust emission and AGB/RSG stars; for three sources in total), and therefore unambiguously identify a third unique component.  
However, any additional separation would not likely grant much improvement in the corrected maps over what we derive here, especially since a third band will trace a slightly different set of (contaminant) sources than those appearing in the 3.6-4.5 $\mu m$ window.  

Moreover, as long as the combined contaminant distribution--which must be more Gaussian than either of the two statistically independent distributions--is itself still significantly different from Gaussian, then it can be identified with ICA, and so can its counterpart (here the corrected image s1-ICA representing the old stars).\footnote{Note that this explicitly assumes that our maps identified as s2-ICA, and not s1-ICA, reflect more than one source, as the preliminary comparisons above would seem to suggest.}  In addition, as discussed below, we find that the detection and removal of contaminants using only two bands has a very minimal effect on the corresponding old stellar map (see $\S$ \ref{sec:dust}).\\
\section{Quality Assessment of ICA separation}
\label{sec:effects}
The two ICA sources, detected via combination of the two input measurement bands at 3.6 and 4.5 $\mu m$, display remarkable consistency with the expected characteristics of the main sources of emission at 3.6 $\mu m$.  Removal of the contaminant emission, which is well correlated with the non-stellar emission imaged at 8 $\mu m$, reveals an underlying old stellar disk with [3.6]-[4.5] colors that are consistent with a population dominated by old K and M giants (\citealt{pahre2}; see Table \ref{tab-colors}).   This is an especially favorable outcome since we expect little to no difference in the colors of stars at these wavelengths, so far along the Rayleigh-Jeans tail of the stellar SED.   The fact that the colors in the final s1 maps are consistent with the colors of late-type giants suggests that each reflects a (nearly) uniformly old stellar population free of contamination from dust and bright intermediate-age stars.  Final source images are not without their imperfections, though, as discussed below.

\subsection{Image imperfections\label{sec:imperfections}}
Imperfections in final source images  become apparent when the number of true sources becomes manifestly incompatible with the set of only two detected with ICA (e.g. when the  `perfect scaling' of contaminants between 3.6 and 4.5 $\mu m$ discussed in $\S$ \ref{sec:dust} breaks down).  
Large variations in the hot dust-to-PAH ratio, in particular, can reduce the effectiveness of the ICA representation of some part of the mapped non-stellar emission and lead to either the over- or under-subtraction of the contaminant emission from the old stellar light.  As described in $\S$ \ref{sec:hotdust}, a contaminant color dominated more by the hot dust than the PAH may result in an underestimation of the flux from PAH away from, e.g., HII regions with the highest contribution from hot dust.  

At least for this small sample any underestimation of fluxes appear modest, and the influence of any error in the contaminant color on the old stellar color even smaller; a change in the contaminant color introduces a change in the old stellar color proportional to the ratio of the contaminant-to-old star flux (see eq. \ref{eq:ica}).  In fact, adjustments to the old stellar color are on the order of, or within, the photometric color errors even with a boost in the contaminant flux by the amount implied by the range in the s2 color (i.e. the uncertainty measure in Table \ref{tab-colors}).

Galaxies with the greatest variation in their hot dust-to-PAH ratio will be the most susceptible to this error.  The more frequent the HII regions, or the more prominent the AGN--i.e. where the hot dust becomes dominant--the greater the risk that the remaining PAH contaminant flux is underestimated.  (Note, though, that the resulting error in the old stellar color is expected to remain small, given the proportionality discussed above.) 

Final source images are also susceptible to the opposite trend, i.e. the overestimation of the hot-dust flux when the PAH dictates the color of the contaminant map.  This overestimation will be limited to small, isolated (HII) regions (at least if PAH is dominant and spatially extensive enough that it does indeed control the s2 color, e.g., in M51). 
Evidence for overestimation in the flux of hot dust is clear in Figure \ref{fig-scat} for M51, for instance, which features a clear upturn at the highest values of $F_{8,ns}$, but in only a small number of pixels.  In this galaxy and in NGC 4321, the overestimation is large enough that there is no remaining flux at the sites of HII regions (cf. \citealt{bendo}) in map s2 (i.e. with overestimation of the hot-dust flux, all emission in the original maps is associated with contaminants). 
Likewise, fluxes of the RSG/AGB stars with colors much in excess of the ICA contaminant color will tend to be overestimated in the contaminant map, occasionally leaving over-subtracted holes in the old stellar light at these locations.  

\indent Although this is a clear loss of information in the old stellar maps in these regions (which would be masked, in any case, in the uncorrected images), the fluxes and colors elsewhere should be more realistic than given a redder ICA contaminant color, since the contaminant flux is overall better representative of the spatially predominant PAH emission.  

Straightforward corrections to the old stellar map at the locations of these over-subtracted holes are possible.  
An obvious option is to replace the holes with fluxes estimated from neighboring pixels, or to use models of the stellar surface brightness (derived with these regions masked) for the substitution.  In both cases, the replacement should be fairer than would be possible, e.g., at the sites of HII regions in the uncorrected images, where neighboring regions may themselves reflect a non-neglible contribution from contaminants.  

These substitutions could then also be used to improve the ICA estimates for either the non-stellar flux in HII regions or the contribution of  intermediate-age stars to regions dominated by these objects, when overestimated in contaminant maps at the location of the old stellar holes.   As a revision to the upper bound on the flux from these components in the contaminant map, an estimate for the flux can be established with the residuals between contaminant-free models and the original 3.6 $\mu m$ emission in these zones.   Alternatively, it is possible to derive a correction factor for, e.g., hot dust fluxes directly from information in the contaminant map (i.e. the color and an underlying level of PAH emission estimated from neighboring regions), together with an assumed [3.6]-[4.5] color or spectral index. 

As the zero-valued holes are limited to small, isolated regions, we argue that this cost is worth the overall high quality with which the old stellar light can be mapped.  In a forthcoming paper, we will expand the investigation of the ICA-identified stellar light distribution initiated below, considering, in particular, the implications for stellar mass maps.  We find that our removal of contaminants enhances 3.6 $\mu m$ images as tracers of the stellar mass surface density.
\subsection{The Old Stellar Light: comparison with NIR\label{sec:oldstars}}
\begin{figure*}[htbp] 
   \centering
   \begin{tabular}{rr}
   \small NGC 1566&\small NGC 2976\\
\includegraphics[width=0.42\textwidth]{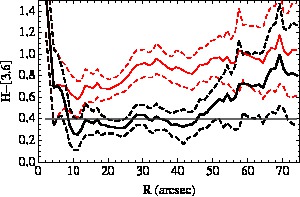}& \includegraphics[width=0.42\textwidth]{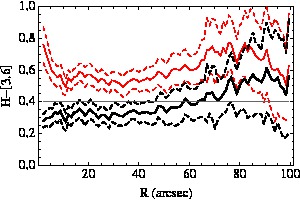}\\
\small NGC 3031 & \small NGC 3184\\
\includegraphics[width=0.42\textwidth]{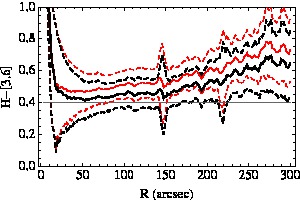} &\includegraphics[width=0.42\textwidth]{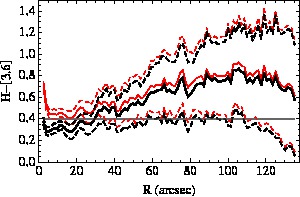}\\
\small NGC 4321&\small NGC 5194\\
\includegraphics[width=0.42\textwidth]{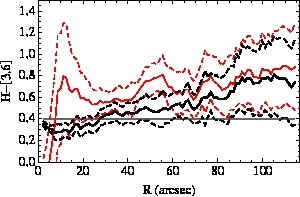} &\includegraphics[width=0.42\textwidth]{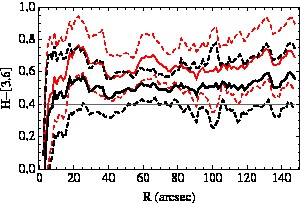} \\
\end{tabular}
   \caption{\small H-[3.6] color profiles for NGC 1566 (top left), NGC 2976 (top right), NGC 3031 (middle left), NGC 3184 (middle right), NGC 4321 (bottom left) and NGC 5194 (bottom right).  The profiles for the original  and corrected maps are shown in thin red and thick black solid lines, respectively, with dashed errors calculated from the range of values at each radius.  The H-[3.6] color expected for an old, dust free population, H-[3.6]=0.4 (see text), is shown as a gray horizontal line.}
   \label{fig-colorprofiles}
\end{figure*}

As a demonstration of the quality with which the corrected fluxes at 3.6 $\mu m$ represent emission from the old stellar disk, here we contrast map s1 (and the uncorrected image) for each galaxy with the H-band-observed distribution of light.   Table \ref{tab-colorsH} lists the zero-order best-fit to the H-[3.6] color profiles plotted in Figure \ref{fig-colorprofiles} for 
the original and corrected maps, both of which have been gaussian-smoothed and sampled at the resolution of the 2MASS H-band image.  (The outer limit of the fit region is determined by the H-band sensitivity.)  

The bluer colors for the contaminant-corrected maps in all cases reflect a reduction in the amount of light relative to the original maps; contaminants globally introduce 0.2 mag on average across the sample.  In 2D maps of the H-[3.6] color this reduction occurs in localized zones, i.e. in the spiral arms, and appears in the 1D profiles in Figure \ref{fig-colorprofiles} as a flattening of deviations from the mean in the strongest cases (e.g. NGC 1566 and NGC 5194).  Notably, with emission from contaminants removed, maps s1 show greater consistency with a prediction from synthesized stellar SED fits to an old, dust-free elliptical for 
H-[3.6]$\sim$0.4,\footnote{\scriptsize see http://web.ipac.caltech.edu/staff/jarrett/\\
\hspace{10cm}irac/calibration/galaxies.html}   
 although this is an imperfectly matched benchmark; the current sample draws almost exclusively from the later Hubble types (i.e. star forming disks with dust emission) and so the H-band image itself may not be free of extinction and emission from young stars.  Star formation history and metallicity will themselves also contribute to real differences from the color expected for an old elliptical galaxy. 

Metallicity may very well be contributing to the H-[3.6] colors here, as evidenced by radial color gradients that are largely positive.  This signature of decreasing metallicity with increasing radius has been previously identified in these bands for ellipticals by  \citet{temi}.  
If, as \citet{temi} argue in this case, enhanced, reprocessed stellar emission appears in NIR and not mid- IR images, this compromises the 
NIR bands as abundance-independent measures of local stellar mass.  
Reddening associated with localized extinction (e.g. in dusty spiral arms) is also arguably evident here.  Even after the contaminants are removed, the profiles for the corrected maps show lingering bumps toward higher H-[3.6].  
These two factors suggest that map s1 may be a cleaner direct tracer of the old stellar disk than the H-band image;     the ICA correction allows 3.6 $\mu m$ imaging to better fulfill its potential as the optimal window on the oldest stars.   But, to be fair, the value of ICA correction is less clear globally or in 1D than in 2D.  
Non-stellar emission contributes only very modestly to the integrated light at 3.6 $\mu m$ and therefore amounts to a small correction in the global H-[3.6] color.  The higher local contribution from contaminants tabulated in Table \ref{tab-contamfraction}, on the other hand, implies as much as 0.45 mag correction to the H-[3.6] color in dusty regions and on average 0.8 mag change to the H-[3.6] color of red clusters.  This clearly suggests that errors in the mass surface density of this size or larger (depending on the M/L) would be introduced with conversion to stellar mass without prior removal of the non-stellar emission.  \\
\begin{table}
\begin{center}
\caption{H-[3.6] colors\label{tab-colorsH}}
\begin{tabular}{rcc}
\tableline\tableline

 Galaxy&Original&s1-ICA\\
\tableline
NGC 1566&0.81$\pm$0.004&0.40$\pm$0.004\\
NGC 2976&0.62$\pm$0.003&0.44$\pm$0.008\\
NGC 3031&0.69$\pm$0.002&0.54$\pm$0.001\\
NGC 3184&0.79$\pm$0.003&0.66$\pm$0.003\\
NGC 4321&0.77$\pm$0.003&0.61$\pm$0.002\\
NGC 5194&0.66$\pm$0.004&0.45$\pm$0.001\\
\tableline
\end{tabular}
\tablecomments{Radially averaged H-[3.6] colors for the original and ICA-corrected images calculated from the profiles shown in Figure \ref{fig-colorprofiles}. (Errors are statistical.)  }
\end{center}
\end{table}
In light of this evidence, ICA-corrected images are favorable to our end goal of mapping 2D stellar mass distributions.  We note that this marks a significant advance in the use of 3.6 $\mu m$ images for this purpose.  The presence of contaminants has typically involved the loss of either 2D or structural (i.e. spiral arm) information, given the use of models (with an axisymmetrically-averaged spiral arm contribution, for example) to replace the 3.6 $\mu m$ emission (i.e. \citealt{deblok}, \citealt{leroy}; \citealt{kendall}; \citealt{kendall1}).  

By comparison, the independence of ICA from any model-inherited assumptions (including those for a particular stellar surface brightness profile, disk shape, or 3.3/8 $\mu m$ PAH ratio) is a keen strength.  
Whereas corrections that rely on models are even susceptible to uncertainties in the assumed center position (e.g. the correction developed by Kendall et al. 2008 that employs a model of the stellar continuum at 8 $\mu m$)--which, moreover, need not be the same between the distributions of light and mass--but no such prior is required to perform ICA.  

Since the ICA correction can maintain full and realistic structural information at 3.6 $\mu m$ (examined in greater detail in Paper II) it is a strong alternative to typical corrections for non-stellar contaminants in this band.  Furthermore, the method is simple and fast, and can be run largely autonomously for a sample as large as S$^4$G.  

\section{Summary and Conclusions}
\label{sec:conclusions}
In this paper we demonstrate the isolation of old stellar light in S$^4$G 3.6 and 4.5 $\mu m$ images from contaminant emission using an ICA technique designed to identify statistically independent source distributions.  ICA decomposes each image at 3.6 and 4.5 $\mu m$ into two other images, tracing either the old stars or the contaminant sources of emission, by maximizing the distinction in the [3.6]-[4.5] color of these components.  We find that the separation achieved with ICA yields a very satisfying view of the old stellar light that, in all cases in the small sample of galaxies analyzed here, has [3.6]-[4.5] colors consistent with the colors of K and M giants, retains a high degree of structural information, and fares as well as, if not better than, other NIR tracers as an optimal tracer of relative or absolute stellar mass.  

The quality of our maps of the old stellar light stems from the ability of ICA to detect all sources of emission apart from the oldest stars: hot dust, PAH and bright but low M/L intermediate-age stars.   By comparing the ICA contaminant maps at 3.6 $\mu m$ to the non-stellar emission imaged at 8 $\mu m$ all three of these sources are clearly identifiable.  In all cases we find that the ICA contaminant map presents a realistic picture of the state of the dust and report 1) the detection of hot dust and 3.3 $\mu m$ PAH emission at  3.6 $\mu m$, which together contribute between 5-13\% to the integrated light of the galaxies in our sample at 3.6 $\mu m$ and on average 20\% locally to star-forming regions and 2) the spatial mapping of regions dominated by intermediate-age AGB or RSG stars, which contribute upwards of 50\% to the 3.6 $\mu m$ flux at the locations of these isolated knots (but less than 5\% to the integrated light of a galaxy).   
 
The hot dust and PAH at 3.6 $\mu m$ together are a potentially useful tracer of the thermal and non-thermal  sources of dust heating in S$^4$G galaxies (i.e. star formation and AGN), following calibration with ancillary 8 $\mu m$, 24 $\mu m$, and/or H$\alpha$ data available for part of the full sample.  The contribution of intermediate-age stars at 3.6 $\mu m$, which is separable from the dust emission via 8$\mu m$ flux (and possibly with uncorrected [3.6]-[4.5] color, as well), may also be useful for probing the diffuse component of 8 $\mu m$ non-stellar emission that does not trace actively star forming regions, but which is instead thought to originate with dust heating by an older population of stars.  
In the future, we will use our maps of the regions dominated by intermediate-age AGB or RSG stars to constrain the fractional contribution of these objects to emission in 3.6 $\mu m$ and blue-ward bands, as well as at 8 and 24 $\mu m$.   

The limit of two input measurement bands enforced here, however, can challenge the detection of these sources with ICA, if only modestly.  The intrinsically reddest regions, i.e. regions dominated by intermediate-age AGB and/or RSG stars, or HII regions where the flux in hot dust exceeds the spatially predominant PAH emission, in particular, tend to be over-subtracted.  Since these are small, isolated zones that would be masked in modelling/structural decompositions in any case, we argue that this is a small cost for the otherwise realistic correction throughout the rest of the contaminated area.  

In addition, we argue that errors associated with the ICA contaminant correction--i.e. the intrinsic incompatibility between the number of real sources and the number of extracted sources--are not significant, the correction even in this case marking an improvement over alternative corrections.   Our use of ICA is superior to, e.g. the scaling and subtraction of the 4.5 $\mu m$ image from the 3.6 $\mu m$ image, or to the subtraction of the non-stellar emission imaged at 8 $\mu m$, 
since it can accommodate for the intrinsically different spectral shapes of the stars and dust between 3.6 $\mu m$ and either 4.5 or 8 $\mu m$.

In this way, our correction for contaminant emission at 3.6 $\mu m$ and/or 4.5 $\mu m$ is optimal for a large data set such as S$^4$G, which samples throughout the Hubble sequence, over a large range of dust contents, gas fractions, and star formation histories.   Furthermore, the correction is self-reliant--using information from only the two IRAC bands available throughout the sample--and can assure uniformity in the treatment of contaminants for all effected galaxies. Contaminants should be identifiable by deviations from the expected colors of K and M giants over significant areas in the original images, i.e. where colors appear in excess of [3.6]-[4.5]=0.   

Corrected 3.6 $\mu m$ images, which supply an optimal view of the underlying old stellar light, are ideal relative tracers of stellar mass and should be indispensable for measuring structural diagnostics such as bar strength and ellipticity, spiral arm-interarm contrasts and relative Fourier amplitudes and phases.   
Our maps can also be converted into absolute mass maps using any of the standard techniques for estimating the stellar M/L (e.g. \citealt{bdJ}; \citealt{z09}), where ancillary optical and/or NIR data is available.  

On the other hand, as will be explored in Paper II, corrected images may support an independent view of the mass distribution in nearby galaxies.   IRAC [3.6]-[4.5] colors, although more sensitive to photometric errors than those defined by a broader wavelength range, are virtually free of the systematic uncertainties plaguing the colors typically used to constrain the stellar M/L, e.g. due to dust extinction/reddening and young stars.   In even the latest generation of maps that actively compensate for these systematics (i.e. \citealt{z09}) there are lingering uncertainties:  convergence on the TP-AGB phase of stellar evolution that dominates NIR light at $\sim$1 Gyr has yet to be reached among different models (cf. Charlot \& Bruzual 2007, \citealt{maraston05} and \citealt{lancon}); where young and/or intermediate-age populations are identified, the M/L can be underestimated given an underlying old stellar population; although dust extinction is accounted for, the reflection and scattering of stellar light by dust is not \citep{z09}.

Leveraging our optimal view of the old stellar light we anticipate generating high-quality 2D mass maps that will provide an enhanced view of structure and mass in nearby S$^4$G galaxies.  Structural decompositions, measurements of bar and/or spiral arm torques, and improved constraints on the baryonic contribution to the distribution of mass in galaxies will be central to the insight on galaxy evolution that S$^4$G can deliver.
\newline
\newline
The authors wish to acknowledge the collective effort of the entire S$^4$G team in this project.   E.A. and A.B. thank the Centre National d'Etudes Spatiales for financial support.  K.S., J.-C.M.-M., T.K. and T.M. acknowledge support from the National Radio Astronomy Observatory, which is a facility of the National Science Foundation operated under cooperative agreement by Associated Universities, Inc.

\end{document}